\documentclass[final,11pt]{article}
\pdfoutput=1
\usepackage{MRnotes}

\newcommand{\slt}{\mathfrak{sl}(2)}
\newcommand{\sut}{\mathfrak{su}(2)}

\newcommand{\Dta}{\mathrm{D}(2,1|\alpha)}

\newcommand{\ts}{\mathfrak{t}}

\newcommand{\zs}{\mathfrak{z}}
\newcommand{\zsb}{\overline{\mathfrak{z}}}
\newcommand{\us}{\mathfrak{u}}

\newcommand{\qb}{\overline{q}}

\newcommand{\hosc}{\Delta_{\mathrm{osc}}}
\newcommand{\hoscb}{\bar{\Delta}_{\text{osc}}}
\newcommand{\josc}{m_{\mathrm{osc}}}
\newcommand{\joscb}{\overline{m}_{\text{osc}}}

\newcommand{\hs}{h_{_\mathrm{CFT}}}
\newcommand{\hsb}{\bar{h}_{_\mathrm{CFT}}}

\newcommand{\asx}[1]{\mathrm{AdS}_3 \times \mathbf{S}^3 \times #1}
\newcommand{\stso}{\mathbf{S}^3 \times \mathbf{S}^1}
\newcommand{\schgS}{\bm{\chi}^\mathrm{g}_{_\mathrm{S}}}
\newcommand{\schgbS}{\overline{\bm{\chi}}^\mathrm{g}_{_\mathrm{S}}}
\newcommand{\schgL}{\bm{\chi}^\mathrm{g}_{_\mathrm{L}}}

\newcommand{\schr}[1]{\bm{\chi}_{_{#1}}}

\renewcommand{\=}{\; = \;}
\renewcommand{\order}[1]{\text{O}(#1)}

\title{Strings and near-extremal black holes in theories with large $\mathcal{N}=4$ superconformal symmetry}
\author{Sameer Murthy${}^{a}$, Mukund Rangamani$^{b}$}

\affiliation[a]{
  Department of Mathematics, King’s College London, The Strand, London WC2R 2LS, UK}
\affiliation[b]{
  Center for Quantum Mathematics and Physics (QMAP)\\
  Department of Physics \& Astronomy, University of California, Davis, CA 95616 USA}

\emailAdd{sameer.murthy@kcl.ac.uk}
\emailAdd{mukund@physics.ucdavis.edu}

\abstract{
We study the one-loop partition function of superstrings in the $\asx{\stso}$ background.
Specifically, we show that the supergravity spectrum, which contains non-chiral primary states unlike other similar AdS$_3$ backgrounds, can be recovered by this partition function in the semiclassical limit. 
We also show how the boundary currents are encoded in the string spectrum. 
Furthermore, we discuss the effect of these boundary currents in the quantum partition function of near-extremal black holes in theories with large $\mathcal{N} = (4,4)$ supersymmetry,  recovering results 
consistent with the analysis of the near-horizon (super-)Schwarzian theory. 
In particular, we show how the BPS index of the large $\mathcal{N} = (4,4)$ theory, which turns out to be temperature dependent, captures the spectrum of excitations around 
supersymmetric BTZ black holes. 
Finally, we comment on the limit when the AdS curvature radius is string scale, which is directly accessible within the RNS formalism for this compactification. 
The spectrum of the limiting theory we obtain differs from the tensionless string spectrum derived in the literature, suggesting that there is not a unique worldsheet CFT for this value of parameters.
}

\begin{document}
\maketitle


\section{Introduction}\label{sec:intro}

Recently, in~\cite{Ferko:2024uxi} we reanalyzed the one-loop partition function for strings on \AdS{3} backgrounds (building on the original results of~\cite{Maldacena:2000kv}), specifically for $\asx{T^4}$ in the case of the superstring. In the present paper, we extend the analysis to superstring compactification on
$\asx{\stso}$ for generic values of \AdS{3} size in string units. This  background preserves \emph{large} $\mathcal{N} =(4,4)$ supersymmetry. The rationale for undertaking this exercise is the following:
\begin{itemize}[wide,left=0pt]
  \item Attempts to understand the holographic AdS/CFT duality in this context~\cite{Boonstra:1998yu,Elitzur:1998mm,deBoer:1999gea,Gukov:2004ym,Gaberdiel:2013vva,Tong:2014yna,Eberhardt:2017fsi,Eberhardt:2017pty,Eberhardt:2019niq,Witten:2024yod} have proven to be subtle despite the large amount of supersymmetry.
  Initially, there appeared to be a tension between the spectrum of excitations derived from the worldsheet and from supergravity. 
  However, following a re-analysis of supergravity spectrum in~\cite{Eberhardt:2017fsi} exact agreement with the canonically quantized string spectrum was demonstrated in~\cite{Eberhardt:2017pty}.    
  A curious feature is that one has  not only short chiral primary multiplets, but also long multiplets among the supergravity states. Obtaining the same spectrum from a one-loop analysis will serve to demonstrate consistency between the Hamiltonian and path integral perspectives.
  \item  The thermodynamics of near-extremal black holes receives quantum corrections from nearly-gapless modes localized in the near-horizon region~\cite{Ghosh:2019rcj,Iliesiu:2020qvm,Heydeman:2020hhw}. The nature of such corrections depends on the supersymmetries of the underlying theory, and those preserved by the black hole. Our analysis allows us to investigate solutions preserving the large $\mathcal{N} =4$ superalgebra, which has not been analyzed hitherto.\footnote{
    The recent work~\cite{Heydeman:2025vcc} analyzes super-Schwarzian theories with a wide range of supersymmetries, including the case of interest here. An explicit discussion of BPS black holes in $\asx{\stso}$ is slated to appear concurrently with our work in~\cite{Heydeman:2025s3s}. We thank Joaquin Turiaci for many insightful discussions in this context. }
\end{itemize}

To set the stage for our analysis, let us recall some salient features of the
$\asx{\stso}$ geometry. Without loss of generality, we will view it as a Type IIB
superstring background, the IIA description being related by T-duality on  the
$\mathbf{S}^1$. The geometry can be analyzed in the RNS formalism when the
compactification is supported by NS-NS B-flux threading the two $\mathbf{S}^3$s.
The worldsheet description comprises $\slt$ and $\sut$ super-WZW models,
along with a free $\mathcal{N}=1$ superconformal field theory.
An advantage of this string compactification is that it allows us to
access the regime of string scale \AdS{3} geometries, including the tensionless
limit in the RNS formalism~\cite{Gaberdiel:2018rqv}, which was recently
exploited in~\cite{Gaberdiel:2024dva}. This observation, in particular, should 
allow one to examine near-extremal BTZ black hole thermodynamics
deep within the stringy regime. This contrasts with the compactifications on
$T^4$ or $K3$ where one is unable to access the tensionless string limit within
the RNS formalism.

We first analyze the worldsheet torus partition function for the target space
having a Euclidean \AdS{3} component with periodically identified time
coordinate. This partition function, which was previously computed in~\cite{Ashok:2020dnc},\footnote{We thank the referee for alerting us to this paper.} upon integration over the worldsheet
modulus, captures the contribution of a single saddle point geometry to the dual
spacetime CFT free energy. This result already suffices to show that the
spectrum of the worldsheet is consistent with the supergravity analysis
of~\cite{Eberhardt:2017fsi}, as was argued earlier from the zero-mode analysis
in~\cite{Eberhardt:2018sce}. Moreover, we demonstrate that the zero-mode part of
the worldsheet spectrum captures accurately the spectrum of the boundary
supergravitons.

Specifically, as described in~\cite{Ferko:2024uxi} we find
that the worldsheet partition function can be organized into representations of
the superisometry algebra of the background geometry, which in the present case
is the algebra $\Dta \oplus \overline{\Dta}$.
Here $\alpha$ parameterizes the relative sizes of the two $\mathbf{S}^3$s, whose
isometries are $\sut$ R-symmetries of this global algebra. Since the worldsheet
partition function is the spacetime CFT free energy, upon exponentiation (which
implements the sum over the free string Fock space) we obtain a piece of the
superconformal characters. To obtain the full superconformal character, one has
to start with this seed and implement an elliptic average. Geometrically, this
corresponds to the statement that the full superconformal character not only
receives contribution from the thermal \AdS{3} geometry with fixed R-symmetry
chemical potentials, but also from configurations which differ by integral
shifts of these chemical potentials. The latter correspond to geometries with
non-trivial gauge potentials, cf.~\cite{Kraus:2006nb}. Carrying out the exercise, we arrive at the expression for the partition function, cleanly expressed in terms of the spacetime superconformal characters. This discussion largely parallels the analysis in~\cite{Ferko:2024uxi}, without any surprises, as long as $\lads > \ell_s$. 

We then turn to the question of what happens in the tensionless limit, obtained by taking $\lads = \ell_s$ (usually referred to as $k =1$, where $k$ is the $\slt$ level). Here we encounter some subtleties, which we outline, and suggest a potential resolution. One novelty about the $\asx{\stso}$ background is that this limit is accessible within the RNS formalism. In the case of $\asx{K3}$ or $\asx{T^4}$ the curvature radii of $\mathbf{S}^3$ and \AdS{3} are equal. For the superstring, once one decouples the fermions, the central charge of the $\sut$ WZW model describing the $\mathbf{S}^3$ shifts down by two units, leading to a non-unitary bosonic $\sut_{-1}$ WZW model for $k = 1$. While one can attempt to make sense of this model, it is efficacious to pass to the hybrid formalism~\cite{Eberhardt:2018ouy}. 

For the $\asx{\stso}$ background, in contrast, the limit where the AdS spacetime is string scale ($k \to 1$) is simply described by a bosonic $\slt_3$ WZW model, a free scalar (corresponding to the $\mathbf{S}^1$), and ten free fermions~\cite{Gaberdiel:2018rqv}. 
The two $\mathbf{S}^3$s effectively decouple after a suitable field redefinition. 
So the worldsheet string can be directly examined in RNS language at~$k=1$.
This was done in~\cite{Gaberdiel:2018rqv}, and more recently in~\cite{Gaberdiel:2024dva}, where it was argued that only the continuous series representations (and their spectral flows) appear in the string spectrum. 
In particular, the discrete series representations of $\slt$ are lifted from the spectrum in this limit. 
This presents a potential paradox since for the boundary (super)gravitons of \AdS{3} arise precisely from these representations (in the sector with no spectral flow). 
The manner in which one avoids this paradox is interesting -- the boundary gravitons are argued to merge into a continuum representation. 
This has also been justified by working with the continuous series characters of $\slt$ theory~\cite{Gaberdiel:2024dva}, 
which are assembled into a modular invariant partition function based on the canonical spectral analysis.  

In the present paper, we analyze the one-loop partition function for the $k=1$ theory in the RNS formalism.
As we explain, in completing the worldsheet modular integral to extract the spectrum, one has to shift an integration contour. In the process, one finds that 
the pole that would correspond to the discrete state containing the vacuum module falls atop the shifted contour, a signal for the merger with the continuum. While this ratifies the earlier observations, it also raises a question about the higher spin currents. Our analysis does not reveal the presence of chiral characters of the form required for the higher spin symmetries. In addition, we also find discrete series states survive in the spectrum at higher spectral flow sectors. 

Based on our analysis in the present paper, we find the following situation: the $k=1$ RNS partition function, which satisfies all the requirements of a GSO projected worldsheet theory, does not agree with the spectrum proposed in~\cite{Gaberdiel:2024dva}. 
We suggest that these two theories, both of which
may be characterized as the bosonic $\slt_3$ WZW  model,\footnote{The shift $k \to k+2$,  originates from decoupling the fermions in the $\mathcal{N}=1$ $\slt_k$ super-WZW model.} are actually two different consistent RNS worldsheet CFTs. 
One of these is obtained from the family of $\slt_k$ WZW models by taking $k \to 3$, while the other would be 
genuinely new.\footnote{Evidence for this new worldsheet CFT has recently appeared in~\cite{Eberhardt:2025sbi}. We thank Lorenz Eberhardt and Matthias Gaberdiel for informing us of their results and sharing a draft of their upcoming work. } 
As an analogy, one could look to non-compact $c=1$ CFTs. Taking the limit of the minimal models from the $c<1$ side one arrives at the Runkel-Watts theory~\cite{Runkel:2001ng} (see also~\cite{Roggenkamp:2003qp}),  a non-compact irrational CFT, which, however, is distinct from the free non-compact boson. For worldsheet strings, one can consider the $c\to 1$ limit of minimal (non-critical) strings, which differs from the $c=1$ string theory (likewise, similar statements apply in the Type 0 superstring construction in the limit $c\to \frac{3}{2}$). 
A similar observation was recently made in~\cite{Chakraborty:2025nlb} who argued that the 
spectrum of the $k=1$ worldsheet theory differs from that of the symmetric orbifold theory, and suggested 
that the two corresponding spacetime CFTs are related by a~$\mathbb{Z}_2$ twist deformation.  

Having understood the worldsheet spectrum, we turn to asking about the partition function of BTZ black holes.  
As Euclidean manifolds, thermal \AdS{3} and BTZ are related -- the target spacetime is the same three-manifold, albeit with different identification of the spatial and temporal circles. 
Therefore, by a boundary S-modular transformation, one can obtain the partition function of the BTZ geometry from the thermal \AdS{3} one. While the computation of the string one-loop determinant around the thermal \AdS{3} background does not by itself lead to a modular invariant spacetime CFT partition function, 
it allows us to access 
the density of states of the dual CFT. 
We will be interested in the grand canonical partition function around the BTZ geometry, with chemical potentials for the R-charges. 

The dominant contribution in the near-extremal BTZ regime can be shown to arise from the boundary supergravitons, whose contribution to the partition function is just the vacuum large $\mathcal{N}=4$ superconformal character. We deduce the low-temperature behavior for both the NS and R sector trace for generic values of the R-charge chemical potentials. The result is shown to be consistent with the expectations from semiclassical gravity for $k>1$. One can already anticipate this result from the spacetime CFT using the general arguments based on the twist gap put forth in~\cite{Ghosh:2019rcj, Iliesiu:2020qvm} for $k>1$. We show that the grand canonical partition function, viz., the NS sector trace, has a power-law suppression by $T^5$ which comes from the 10 bosonic near-horizon gapless modes. 
This translates in the canonical ensemble with fixed charges to the now-familiar $T^{3/2}$ suppression due to the near-horizon Schwarzian modes. 
However, one also has a further exponential damping because the states that contribute have non-vanishing energy, so $\mathscr{Z}_\mathrm{NS} \sim T^{3/2}\, e^{-\beta\, E_Q}$, where $E_Q$ is the energy carried by a charge quantum. 

One can similarly consider the partition function with periodic boundary conditions for the fermions, leading to the Ramond sector trace around the BTZ geometry. The grand canonical ensemble partition function
can be used to extract the density of states with fixed energy and charges.  There is still a non-vanishing charge activation energy, and so the result is still exponentially damped with a different power law prefactor. 

The more interesting case is when the R-charge chemical potentials are tuned to obtain the index of the large $\mathcal{N}=4$ algebra defined in~\cite{Gukov:2004fh} (see also~\cite{Gukov:2004ym}). 
This index zeros in on the short multiplets. Since the algebra has fermionic generators apart from the supercharges, obtaining a non-zero index requires soaking up additional fermion zero mode. 
One therefore should compute the analog of the
``new supersymmetric index" in 2d SCFTs~\cite{Cecotti:1992qh} or the helicity supertrace studied in the context of flat space black holes.
At the same time, one should also remove the bosonic zero mode  associated with the $\mathbf{S}^1$ translations, which can be done by working in a fixed charge sector. 
The appearance of these fermionic and bosonic zero modes is similar to  the case of~$\asx{T^4}$.
Upon removing these fermionic and bosonic zero modes, one finds an answer that is not holomorphic. 
Specifically, in the canonical ensemble, we find that this index is exponentially damped. While this may seem unexpected, it is consistent with unitarity bounds in the CFT. In theories with large $\mathcal{N}=4$ symmetry, the Ramond sector BPS states do not lie at energy $E = \frac{c}{12}$ -- states have an energy that scales quadratically with the R-charge above this value. The index we compute picks up this excess energy and gives a prediction for its spectral distribution.

The outline of the paper is as follows. We begin in~\cref{sec:sugra} with a review of the spectrum of the dual spacetime CFT obtained from supergravity. We then turn to the large $\mathcal{N}=4$ superconformal algebra in~\cref{sec:N4algebra}. Specifically, we review its basic features, highlighting some discussion from prior literature on the BPS representations and indices. Armed with these preparations, we proceed in~\cref{sec:wsstring} to outline the determination of the one-loop string partition function. We show  that this one-loop analysis reproduces the supergravity spectrum in the semiclassical limit. In the stringy regime, one still has a twist gap separating the vacuum multiplet from the rest of the spectrum, as long as the curvature radius of AdS is above the string scale ($k>1$). We analyze  the partition function in the $k=1$ theory and point out  subtleties in relating the resulting spectrum with that of the tensionless string. 
In~\cref{sec:nextBTZ} we then turn to the near-extremal BTZ black hole spacetime, whose low-temperature partition function we derive from our worldsheet answer. We specifically argue that after soaking up the additional zero modes of the $\mathbf{S}^1$ superisometry, we recover the index counting the BPS degeneracy. The  appendices contain some technical results:~\cref{sec:N4algcomm} reviews the large $\mathcal{N} =4$ algebra, while~\cref{sec:tensionless} contains details of the tensionless string RNS partition function calculation, and ~\cref{sec:Rsums} deals with evaluating the low-temperature limit of the partition function.  

\section{Supergravity on \texorpdfstring{$\asx{\stso}$}{AdS3S3S3S1}}\label{sec:sugra}

The starting point of our discussion is the $\asx{\stso}$ geometry supported by
NS-NS flux. The geometry is built from two sets of NS5-branes -- $k_+$ of which
wrap one $\mathbf{S}^3$ and $k_-$ wrap the other $\mathbf{S}^3$. Both of these sets also wrap the $\mathbf{S}^1$. There are in addition $N_1$ fundamental strings. In the three-dimensional space transverse to the compact directions, we have a set of effective strings, whose near-horizon is the \AdS{3} geometry. The target space line element which solves the low-energy supergravity equations of motion is (cf.~\cite{Boonstra:1998yu,Gukov:2004ym} for a detailed discussion of the supergravity solutions)
\begin{equation}\label{eq:dssq}
\begin{split}
ds_{10}^2
\=
\lads^2 \left(d\rho^2 + \cosh^2\rho\, d\tE^2 + \sinh^2\rho \,
d\varphi^2 \right) + R_+^2\, d\Omega_+^2 + R_-^2\, d\Omega_-^2  + L^2\, dy^2
\,.
\end{split}
\end{equation}
Here $d\Omega_\pm^2$ is the line element on a unit round $\mathbf{S}^3$.  The 
geometry is supported by the NS-NS three-form,
\begin{equation}\label{eq:Hflux}
\begin{split}
H
\=
\frac{2}{\lads} \, \omega_\mathrm{AdS} + \frac{2}{R_+}\, \omega_+ +
\frac{2}{R_-}\, \omega_- \,,
\end{split}
\end{equation}
where $\omega_\text{AdS}$ and $\omega_\pm$ are the volume-forms of spaces with
unit radii.
The configuration is characterized by four length scales $\{\lads, R_+, R_-, L\}$  corresponding to each of the four factors. These are determined by three integers~$(k_+,k_-,N_1)$, 
which correspond to the quantized fluxes of NS5-branes and fundamental strings. 
They are related to the parameters of the solution as
\begin{equation}\label{eq:scales}
\frac{R_\pm}{\ell_s}
 \= \sqrt{k_\pm} \,,
\qquad
\frac{L}{\ell_s} \= g_s^2\, \frac{N_1}{k_+\, k_-\,\sqrt{k_+ + k_-}}\,.
\end{equation}

This target space geometry is a viable string background in both Type IIA and
IIB string theory, and preserves a
superisometry group $\Dta \times \overline{\Dta}$ in the bulk spacetime. There is additionally a $\mathrm{U}(1)$ global symmetry associated with the $\mathbf{S}^1$, which remains decoupled at the level of supergravity. The
parameter $\alpha$  refers to the ratio of the radii of the two $\mathbf{S}^3$
\begin{equation}\label{eq:alphadef}
\alpha \=  \frac{k_-}{k_+}\,.
\end{equation}
The superalgebra has two bosonic $\sut$ subalgebras, which correspond to the $R$-symmetry. Finally, there is a $\slt$ global conformal algebra and the superpartners of these bosonic isometries.

Due to the \AdS{3} asymptotics, the superisometry algebra gets enhanced asymptotically to two 
copies of the large $\mathcal{N} =4$ superconformal algebra,  $\mathcal{A}_\gamma \oplus \overline{\mathcal{A}}_{\gamma} $. This algebra comprises two  $\sut$ current algebras at level $\mathsf{k}_\pm = N_1\, k_\pm$, along with the super-Virasoro generators.\footnote{ We indicate the levels $\mathsf{k}_\pm$ of the spacetime ($\mathcal{N}=4$) CFT superconformal symmetry with a  sans serif font, reserving $k_\pm$ etc., for the worldsheet ($\mathcal{N}=1$) current algebra levels. } 
The parameter $\gamma$ is 
related to $\alpha$ as
\begin{equation}\label{eq:gammadef}
\gamma \= \frac{\mathsf{k}_-}{\mathsf{k}_+ + \mathsf{k}_-}  
\= \frac{\alpha}{1+\alpha} \,.
\end{equation}
In terms of these quantities, the AdS radius in string units is given by
\begin{equation}
    \frac{\lads}{\ell_s} \= \sqrt{k} \,.
\end{equation}
The spacetime CFT central charge is given by the AdS radius in Planck units~\cite{Brown:1986nw} and can be expressed as 
\begin{equation}\label{eq:ckdef}
c \= 6\,k \, N_1 \,\equiv\,  6\, \mathsf{k}\,, \qquad
k \,\equiv\,\   \frac{k_+ \,k_-}{k_+ + k_-} \,,
\end{equation}

As indicated above, the parameter $k$ is the harmonic sum of the levels $k_\pm$ of the two $\sut$ subalgebras.  However, this superconformal algebra has a $\mathfrak{u}(1)$ current algebra (supplemented with 4 fermionic generators) as a subalgebra. Geometrically, the $\mathfrak{u}(1)$ generators arise from the $\mathbf{S}^1$. While this component was decoupled at the level of supergravity, in the full superconformal algebra one encounters non-trivial commutators. As an aside, as we briefly review in~\cref{sec:N4algcomm}, by a field redefinition~\cite{Goddard:1988wv} one can decouple the $\mathfrak{u}(1)$ current algebra and four free fermions from $\mathcal{A}_\gamma$. The decoupled algebra $\widetilde{\mathcal{A}}_\gamma$ has a central charge $c-3$. The $\mathfrak{u}(1)$ current algebra and fermions carry the remaining central charge and have a free field realization.

\subsection{Spacetime supercharacters}\label{sec:sisochar}

Let us first record some useful results relating to the superisometry algebra (we describe an analogous result for the superconformal algebra in~\cref{sec:sconfchar}). We will express our results for the string
partition function directly in terms of the characters we introduce below.

\paragraph{Representations of the superisometry algebra:}
Since $\Dta$ has two $\sut$ subalgebras, its multiplets are labeled by a conformal weight $h$ and $\sut_\pm$ spins $\ell_\pm$. Unitarity demands that a generic multiplet in the NS-sector satisfy\footnote{ All the unitarity bounds we present should be viewed in terms of the representation labels. We use $\ell_\pm$ for the most part to denote the representation and not to refer to the $J_3^\pm$ eigenvalues of the states. }
\begin{equation}\label{eq:dtauni}
h \;\geq\; \gamma \, \ell_+ + (1-\gamma)\, \ell_- \,.
\end{equation}
The representation content of a generic multiplet can be summarized by presenting the $\Dta$ character, which is 
\begin{align}\label{eq:dtalong}
\schgL(h,\ell_+,\ell_-)
 & \= 
\bigl(q^h+ q^{h+2} \bigr)\, \chi_{\ell_+}(r_+)\, \chi_{\ell_-}(r_-) 
\nonumber \\
 & \qquad
+ \left( q^{h+\frac{1}{2}} + q^{h+\frac{3}{2}}\right) \bigg(
\chi_{\ell_+ +\frac{1}{2}}(r_+) +  \chi_{\ell_+ -\frac{1}{2}}(r_+) \bigg)
\bigg( \chi_{\ell_- +\frac{1}{2}}(r_-)+  \chi_{\ell_- -\frac{1}{2}}(r_-)  \bigg)
\nonumber \\
 & \qquad
+q^{h+1} \bigg(2\,\chi_{\ell_+}(r_+)\, \chi_{\ell_-}(r_-) + \Bigl(\chi_{\ell_+ + 1}(r_+) + \chi_{\ell_+ - 1}(r_+) \Bigr) \chi_{\ell_-}(r_-)  \nonumber \\
 & \qquad \qquad\qquad \qquad
+\chi_{\ell_+}(r_+) \Bigl(\chi_{\ell_- + 1}(r_-) + \chi_{\ell_- - 1}(r_-) \Bigr)   \bigg) \,.
\end{align}
We have written this expression using the standard $\sut$ characters
\begin{equation}\label{eq:su2char}
\chi_\ell(r) \=
\frac{r^{\ell+ \frac{1}{2}} - r^{-\ell- \frac{1}{2}}}{r^{\frac{1}{2}}- r^{- \frac{1}{2}}} \,, \qquad \ell \geq 0 \qquad \qquad ( \text{and} \; \; \chi_\ell(r)= 0 \,, \; \;  \ell <0) \,,
\end{equation}
and have used
\begin{equation}\label{eq:qrpmdef}
q = e^{2\pi i\,\tau} \,, \qquad r_\pm  = e^{2\pi i\, \rho_\pm} \,.
\end{equation}
We reserve $q$ and $\tau$ to denote the complex structure data of the \AdS{3} boundary, and introduce different symbols for the  worldsheet modular parameter later. 

Saturation of the inequality in~\eqref{eq:dtauni} results in a short multiplet, whose representation content is instead given by
\begin{equation}\label{eq:dtashort}
\begin{aligned}
\schgS(\ell_+,\ell_-)
 & \=
q^{h_s}\,  \chi_{\ell_+}(r_+)\, \chi_{\ell_-}(r_-)
+ q^{\frac{3}{2}}\, \chi_{\ell_+ - \frac{1}{2}}(r_+) \,
\chi_{\ell_- -\frac{1}{2}}(r_-)                   \\
 & \quad
+ q^{h_s+\frac{1}{2}} \bigg(
\chi_{\ell_+ -\frac{1}{2}}(r_+)\, \chi_{\ell_- -\frac{1}{2}}(r_-)
+ \chi_{\ell_+ +\frac{1}{2}}(r_+)\, \chi_{\ell_- -\frac{1}{2}}(r_-)
+ \chi_{\ell_+ -\frac{1}{2}}(r_+)\, \chi_{\ell_- -\frac{1}{2}}(r_-)  \bigg) \\
 & \quad
+ q^{h_s+1} \bigg(
\chi_{\ell_+}(r_+)\, \chi_{\ell_- - 1}(r_-)
+ \chi_{\ell_+ -1 }(r_+)\, \chi_{\ell_-}(r_-)
+ \chi_{\ell_+}(r_+)\, \chi_{\ell_-}(r_-)  \bigg) 
\,,
\end{aligned}
\end{equation}
with $h_s$ denoting the BPS weight 
\begin{equation} \label{eq:defhs}
h_s = h_s(\ell_+,\ell_-) \equiv \gamma\, \ell_+ + (1-\gamma)\, \ell_-   \,.
\end{equation}

Given this, it is not difficult to see that the  long multiplet character can be written as the combination of characters of two short multiplets with different spins:
\begin{equation}\label{eq:DtaLS}
\begin{split}
\schgL(h,\ell_+,\ell_-)
 & \=
q^{h-h_s(\ell_+,\ell_-)} \Bigl[ \schgS(\ell_+,\ell_-) +  \schgS\left(\ell_{+}+ \frac{1}{2},\ell_- + \frac{1}{2}\right) \Bigr] \,.
\end{split}
\end{equation}
This identity holds for generic long multiplets.

There are truncations in the short multiplet character at small values of the spin. In particular, for $\ell_+ = \ell_- = \frac{1}{2}$ we find
\begin{equation}\label{eq:dtalowspin}
\schgS\left(\frac{1}{2},\frac{1}{2},\frac{1}{2}\right)
\=
-q^{\frac{1}{2}} \, \chi_{\frac{1}{2}}(r_+)\, \chi_{\frac{1}{2}}(r_-)
+ q\left( 1+ \chi_{1}(r_+) + \chi_1(r_-)\right) - q^{\frac{3}{2}} \, \chi_{\frac{1}{2}}(r_+)\, \chi_{\frac{1}{2}}(r_-) + q^2\,.
\end{equation}
These characters will be useful for us since the single string contribution to the spacetime partition function will naturally be expressed in terms of them.

\subsection{Supergravity spectrum}\label{sec:sugraspectrum}

The spectrum of supergravity fields compactified on $\asx{\stso}$ was analyzed thoroughly in~\cite{Eberhardt:2017fsi}. The authors carried out a harmonic decomposition of Type IIB supergravity fields on $\mathbf{S}^3 \times \mathbf{S}^3 \times \mathbf{S}^1$, and organized the resulting fields into $\Dta$ multiplets. They found that the spectrum takes the form
\begin{equation}
\begin{split}
 & \sum_{\ell  =0}^\infty\,
\abs{ \schgS(\ell,\ell) + \schgS\pqty{\ell +\frac{1}{2},\ell+ \frac{1}{2}} }^2
+ \sum_{\ell_+ \neq \ell_-} \, \abs{\schgL\pqty{h, \ell_+, \ell_-}}^2\,.
\end{split}
\end{equation}
The conformal dimension of the short multiplets is fixed by the $\sut$ quantum numbers, as in~\eqref{eq:defhs}.
Further, for the long multiplets the dimension is also given in terms of the spins as\footnote{We are exploiting the fact that $\frac{k_\pm}{k} = \frac{\mathsf{k}_\pm}{\mathsf{k}}$ to write the expression in terms of the levels that will appear in the worldsheet analysis.}
\begin{equation}\label{eq:sugralong}
h(\ell_+, \ell_-) = \sqrt{
  \frac{k}{k_+}\, \pqty{\ell_+ + \frac{1}{2}}^2+
  \frac{k}{k_-}\,\pqty{\ell_- + \frac{1}{2}}^2} - \frac{1}{2}
\end{equation}
Importantly, short multiplets only arise with $\ell_+ = \ell_-$. As remarked above, these are the only ones that can be uplifted into short multiplets of the full superconformal algebra.

Since the conformal dimensions given by~\eqref{eq:sugralong} saturate the
unitarity bound when $\ell_+  = \ell_-$, we can exploit~\eqref{eq:DtaLS} to
present the spectral data in a compact form. The full supergravity spectrum is then simply captured by,
\begin{equation}\label{eq:sugraalt}
\sum_{\ell_+, \ell_-} \, \abs{\schgL\pqty{h(\ell_+,\ell_-), \ell_+, \ell_-}}^2 \,.
\end{equation}
We will reproduce this spectrum from the worldsheet partition function. Note that
in supergravity the spins are unbounded from above, which will fail to be true
once we make the \AdS{} scale finite in string units.

\section{The large \texorpdfstring{$\mathcal{N} =4$}{N=4} 
superconformal algebra and BPS representations}\label{sec:N4algebra}

We now turn to the asymptotic symmetry algebra in \AdS{3}, which extends the superisometry algebra into a superconformal algebra~\cite{Henneaux:1999ib}. 
While the background still preserves 8 supercharges, owing to the presence of the two $\sut$ algebras 
one ends up with the \emph{large} $\mathcal{N}=4$ superconformal algebra. 
We summarize some salient facts about this algebra below (the commutation relations are recorded in~\cref{sec:N4algcomm}). 
Representation theory of this algebra was analyzed in~\cite{Gunaydin:1988re,Petersen:1989zz,Petersen:1989pp} originally, and a nice summary can be found in~\cite{Gukov:2004fh}. We review some salient facts relevant for our analysis below.

Let us begin by noting an important subtlety regarding the representations of the superconformal algebra. The unitarity bound on the multiplet, which is now labeled by the conformal dimension $h$, $\sut_\pm$ spins $\ell_\pm$, and the $\mathfrak{u}(1)$ charge $u$, is different from that of the global algebra. 
Not only is the presence of the $\mathfrak{u}(1)$ charge a new wrinkle, but there are also non-linear contributions from the spin. In contrast to~\eqref{eq:dtauni}, the unitarity bound (for NS sector states) takes the form,  \\
\begin{equation}\label{eq:N4uni}
\begin{split}
h & \; \geq \; \frac{1}{\mathsf{k}_+ + \mathsf{k}_-}\, \bqty{ \mathsf{k}_+\, \ell_- + \mathsf{k}_- \, \ell_+  + (\ell_+ - \ell_-)^2 + u^2 } \\
& \=
 \gamma\, \ell_+ + (1-\gamma) \, \ell_-  + \frac{(\ell_+ - \ell_-)^2 + u^2}{(k_+ + k_-)\,N_1 }  \,.
\end{split}
\end{equation}
The $\sut$ spins are constrained to lie in the window $ 0 \leq \ell_\pm \leq \frac{1}{2} (\mathsf{k}_\pm -1)$.  For $\ell_\pm \in \{0,\frac{1}{2}\}$ the
additional shortening conditions on the multiplet modify these relations. 
The short or BPS representations of the superconformal algebra saturate the unitary bound,
just like the short or BPS representations of the global algebra saturate the bound~\eqref{eq:dtauni}. 
It is essential to note, however, that 
not all short multiplets of the global algebra uplift to short superconformal multiplets. 
It is clear that the only ones that can do so carry no $\mathfrak{u}(1)$ charge and have their $\sut_\pm$ spins equal.

Notice that the non-linear terms in the bound~\eqref{eq:N4uni} involve $N_1 \sim L\,g_s^{-2}$, cf.~\eqref{eq:scales}. 
In particular, for states with $\ell_+ = \ell_-$ and $u=0$ this bound for the superconformal algebra coincides with the unitarity bound~\eqref{eq:dtauni} for the global isometry algebra. 
For such states, the unitarity bound does not depend on  $g_s$ and is a homogeneous function of degree zero in the levels~$k_\pm$ of the current algebras on the worldsheet. 
One can make some useful inferences from this observation. 
Since the supergravity spectrum is insensitive to $g_s$, 
the only BPS states that occur are those with 
$\ell_+ = \ell_-$ and $u=0$. 
Such states saturate the unitarity bound for the global as well as for the full superconformal algebra. 
One can also have non-BPS states in supergravity with $\ell_+ \neq \ell_-$, which respect the bound~\eqref{eq:N4uni} but do not saturate it.
Finally, supergravity states cannot saturate the unitarity bound~\eqref{eq:dtauni} of the global superisometry algebra whilst having $\ell_+ \neq \ell_-$. 
Indeed, since the correction at $\order{g_s^2}$ is positive, 
such states will not respect the bound~\eqref{eq:N4uni} and hence 
fail to be unitary with respect to the full superconformal algebra in the full theory. 
Since BPS states of supergravity obey the bound imposed by the global supersymmetry algebra, they must have~$\ell_+ = \ell_-$. 

As noted earlier, the harmonic analysis on the $\asx{\stso}$ background carried out in~\cite{Eberhardt:2017fsi} fleshes out these observations. 
It demonstrates that the supergravity spectrum has both BPS states, which fall in short multiplets of the global algebra, as well as non-BPS states. 
Whilst the latter is  unusual -- most Freund-Rubin compactifications only result in chiral primary states being retained in the supergravity limit -- the aforementioned observations make clear that this should have been anticipated 
assuming no constraints on the possible $\sut$ spins that could appear in the spectrum. 

It is also useful to examine the R sector unitarity bound, which takes the form 
\begin{equation}\label{eq:Runitarity}
\begin{split}
h - \frac{c}{24} 
&\; \geq\;  
    \frac{1}{\mathsf{k}_+ + \mathsf{k}_-} \bqty{ \pqty{\ell_+ + \ell_- -\frac{1}{2}}^2 + u^2 } \\
&  \= 
    \frac{1}{N_1\,(k_+ + k_-)} \bqty{ \pqty{\ell_+ + \ell_- -\frac{1}{2}}^2 + u^2 }\,.
\end{split}
\end{equation}
The $\sut_\pm$ spins are now constrained to lie in $\frac{1}{2} \leq \ell_\pm \leq \frac{1}{2}\, \mathsf{k}_\pm$. 
For states with $u=0$ this excess energy can be expressed as 
\begin{equation}\label{eq:Runitc}
h - \frac{c}{24} 
\; \geq\;  \frac{6}{c}\, \frac{\alpha}{(1+\alpha)^2} \, \pqty{\ell_+ + \ell_- -\frac{1}{2}}^2\,.
\end{equation}
Consequently, we see that there are no states at the threshold energy $E = c/12$, which in other \AdS{3} examples would have corresponded to BPS black holes. This point will be important when we analyze the spectrum of Ramond states, which should include BTZ black holes. We will return to this after we discuss the characters, BPS spectrum, and the notion of supersymmetric indices for this large $\mathcal{N} =4$ algebra. Since these energy shifts are again of $\order{g_s^2}$ they are invisible in supergravity. However, they could be relevant for the one-loop partition function, which, as we shall see, will indeed be the case.

\subsection{Representations and superconformal characters}\label{sec:sconfchar}

The representations and characters of the large $\mathcal{N} =4$ superconformal algebra were analyzed in~\cite{Petersen:1989zz,Petersen:1989pp}.
Since we will analyze the spectrum of fluctuations around \AdS{3} vacuum, we will primarily be interested in the NS sector character. To allow for analyzing the spectrum in different charge sectors, we will include $\sut_\pm$ chemical potentials and also consider the insertion of~$(-1)^F$. 
The character is defined as a trace over a given representation of the algebra, with $\delta \in \{0,1\}$, 
\begin{equation}\label{eq:N44vacNS}
\schr{\text{NS}}^\gamma(q,r_+, r_-;\delta)
\=
\text{Tr}_{_\mathrm{NS}} \, \pqty{e^{i\pi\, \delta\, F}\, q^{L_0 - \frac{c}{24}}\, r_+^{J_0^{3,+}}\, r_-^{J_0^{3,-}} }\,.
\end{equation}
Here, we have kept the dependence on the representation implicit.

The characters of the long and short multiplets built atop  the state $\ket{h,\ell_+,\ell_-}$ were obtained in~\cite{Petersen:1989zz} and ~\cite{Petersen:1989pp}, respectively. They can be expressed as (cf.~\eqref{eq:qrpmdef} for our conventions)
\begin{equation}\label{eq:N4NSexp}
\begin{split}
\schr{\mathrm{NS}}^\gamma(q,r_+, r_-; \delta)
 & \=  q^{h-\frac{c}{24}} \,
\frac{\vartheta_{3+\delta}^2 \left(\frac{\rho_+ + \rho_-}{2},\tau \right)\, \vartheta_{3+\delta}^2\left(\frac{\rho_+ - \rho_-}{2},\tau\right)}{\eta(\tau)^6\,\vartheta_1(\rho_+,\tau)\, \vartheta_1(\rho_-,\tau) }   \; \mu_{_\mathrm{NS}}(q,r_+,r_-; \delta) \,.
\end{split}
\end{equation}
The theta functions arise from the one-loop determinant, and can be assembled in supergravity by utilizing the heat kernels results of~\cite{Giombi:2008vd,David:2009xg}. 
The information about the nature of the multiplet is in the function $\mu$  which itself takes the form
\begin{equation}\label{eq:mudef} 
\begin{split}
\mu_{_\mathrm{NS}}(q,r_+, r_- ; \delta)
 & \=
\sum_{n,m \in \mathbb{Z}} \, q^{n^2 \, \mathsf{k}_+  + m^2\, \mathsf{k}_- +\, n\, (2\,\ell_+ +1) +\, m\, (2\, \ell_- +1 )} \, \sum_{\epsilon_\pm \,= \, \pm }\, \epsilon_+\, \epsilon_- \, \widehat{\mu}_{_\mathrm{NS}}\left(q,r_+^{\epsilon_+},r_-^{\epsilon_-}\right) \\
\widehat{\mu}_{_\mathrm{NS}}\left(q,r_+ ,r_- \right)
 & =
r_+^{n\,\mathsf{k}_+ + \ell_+ + \frac{1}{2}}\,  r_-^{m\,\mathsf{k}_- + \ell_- + \frac{1}{2}}\,  \times
\begin{cases}
1\,, & \qquad \text{long} \,, \\
\dfrac{1}{1 + e^{i\pi \delta}\, r_+^{\frac{1}{2}}\, r_-^{\frac{1}{2}}\, q^{m+n+ \frac{1}{2}}} \,,
     & \qquad \text{short}\,.
\end{cases}
\end{split}
\end{equation}
In the above formulas, we have used the spacetime CFT levels
$\mathsf{k}_\pm$, with the central charge defined in terms of their harmonic sum
$\mathsf{k}$ as in~\eqref{eq:ckdef}.
In~\cref{sec:wsstring} we will recover these characters from the multi-string free energy around the $\asx{\mathbf{S}^3 \times \mathbf{S}^1}$ geometry.

For completeness, let us also record the result for the Ramond sector characters, even though we won't directly be able to obtain them from the string path integral. These are defined by the analog of~\eqref{eq:N4NSexp} with the trace taken in the R sector. Explicitly, 
\begin{equation}\label{eq:N4Rexp}
\begin{split}
\schr{\mathrm{R}}^\gamma(q,r_+, r_-; \delta)
 & \=  q^{h-\frac{c}{24}} \,
\frac{\vartheta_{2-\delta}^2 \left(\frac{\rho_+ + \rho_-}{2},\tau \right)\, \vartheta_{2-\delta}^2\left(\frac{\rho_+ - \rho_-}{2},\tau\right)}{\eta(\tau)^6\,\vartheta_1(\rho_+,\tau)\, \vartheta_1(\rho_-,\tau) }   \; \mu_{_\mathrm{R}}(q,r_+,r_-; \delta) \,,
\end{split}
\end{equation}
where 
\begin{equation}\label{eq:Rmudef} 
\begin{split}
\mu_{_\mathrm{R}}(q,r_+, r_- ; \delta)
 & \=
\sum_{n,m \in \mathbb{Z}} \, q^{n^2 \, \mathsf{k}_+  + m^2\, \mathsf{k}_- + 2\,\ell_+ \,n + 
2\, \, \ell_- \,m} \, \sum_{\epsilon_\pm \,= \, \pm }\, \epsilon_+\, \epsilon_- \; \widehat{\mu}_{_\mathrm{R}}\left(q,r_+^{\epsilon_+},r_-^{\epsilon_-}\right) \\
\widehat{\mu}_{_\mathrm{R}}\left(q,r_+ ,r_- \right)
 & =
r_+^{n\,\mathsf{k}_+ + \ell_+ }\,  r_-^{m\,\mathsf{k}_- + \ell_- }\,  \times
\begin{cases}
1\,, & \qquad \text{long} \,, \\
\dfrac{1}{r_+^{-1}\,q^{-n} + e^{i\pi \delta}\, r_-^{-1}\, q^{-m}} \,,
     & \qquad \text{short}\,.
\end{cases}
\end{split}
\end{equation}
%

\subsection{BPS states and supersymmetric indices}\label{sec:BPSN4}

The large~$\mathcal{N}=4$ superconformal algebra is discussed in \cref{sec:N4algcomm}.
As mentioned there, the global superisometry algebra~$\Dta$ 
is generated by~$\{L_0, L_{\pm 1}\}$,
$G^a_{\pm \frac{1}{2}}$ and $J_0^{\pm, i}$ in the NS sector. 
The following generators,  
\begin{equation} \label{eq:Neq2subalg}
L_n \,, \quad 
\mathcal{G}^+_r \;\equiv \; G^3_r+i \,G^0_r \,, \quad 
\mathcal{G}^-_r \; \equiv \; G^3_r-i \,G^0_r \,, \quad 
J_n \; \equiv \;  2\,\bigl( \gamma \,J_n^{+,3} - (1-\gamma) \,J_n^{-,3} \bigr)\,,
\end{equation}
define an~$\mathcal{N}=2$ superconformal subalgebra~\cite{Gukov:2004fh, Gaberdiel:2013vva} (we  recall that~$\gamma = \frac{\alpha}{1+\alpha} = \frac{k_-}{k_++k_-}$ and assume $\alpha \neq 1$).
The modes obey the following commutation relations, 
\begin{equation}
\begin{split}
     \{ \mathcal{G}^+_r,\mathcal{G}^-_s \}  &\= 4\, L_{r+s} + 2\,(r-s) \,J_{r+s} + \frac{c}{3} \, \delta_{r+s,0} \, \bigl(r^2-\tfrac14 \bigr) \,, \\    
     [J_n,\mathcal{G}^{\pm}_r]  & \= \pm \mathcal{G}^{\pm}_{r+n} \,.
\end{split} 
\end{equation}
The global part of this subalgebra contains the commutators
\begin{equation}
     \{ \mathcal{G}^+_{-\frac12},\mathcal{G}^-_{\frac12} \}  \= 4 L_0 - 2 J_0\,. 
\end{equation}
Chiral primary (BPS) states are annihilated by~$\mathcal{G}^+_{-\frac12},\mathcal{G}^-_{\frac12}$ and hence obey~$2L_0=J_0$. 

The elliptic genus defined with respect to the above~$\mathcal{N}=2$ subalgebra is the following trace in the left-moving sector, 
\begin{equation}  \label{eq:ellgenNS}
    \text{Tr}_{_\mathrm{NS}} \pqty{ (-1)^{F_s} \, q^{L_0-J_0/2} }\,,
\end{equation}
and it only receives contributions from the above BPS states.  
This index can be expressed in terms of the 
traces of the large~$\mathcal{N}=4$ algebra of the  type~\eqref{eq:N44vacNS} 
upon setting~$\rho_+=\gamma \,\tau$, $\rho_-=-(1-\gamma)\,\tau$, so that~$\rho_+-\rho_-=\tau$. 
We use this observation in~\cref{sec:nextBTZ} below. 
To be more precise, the trace~\eqref{eq:ellgenNS} actually vanishes on both long and short representations of the large~$\mathcal{N}=4$ algebra. 
This can be seen explicitly from the characters~\eqref{eq:N4NSexp}. 
When we set~$\rho_+-\rho_-=\tau$, the characters of the long representations in the NS-sector have a double zero from the zeros of the theta functions.
The characters of the short representations have a single zero because of the presence of the simple pole in the function~$\widehat \mu$ given in~\eqref{eq:mudef}.

One can define an index directly in the  chiral sector of the large~$\mathcal{N}=4$ algebra as
\begin{equation} \label{eq:largeN4index}
\Bigl(\frac{d}{d\rho_+}-\frac{d}{d\rho_-} \Bigr)
\text{Tr}_{_{\mathrm{NS}}} \pqty{ (-1)^{F_s} \, q^{L_0 - \frac{c}{24}}\, r_+^{J_0^{3,+}}\, r_-^{J_0^{3,-}} }\biggr|_{\rho_+ \,=\, \gamma \tau \,, \; 
\rho_- \,=\, -(1-\gamma)\tau} \,.
\end{equation}
This index only receives contributions from the short representations, and is equivalent to the elliptic genus with the insertion of the fermionic zero mode as in~\cite{Cecotti:1992qh, Maldacena:1999bp}. 

From the physical perspective, we understand the above index as follows. 
Recall that the~$A_\gamma$ algebra has a bosonic zero mode, corresponding to the translations along the~$\mathrm{U}(1)$ symmetry direction. 
The superpartner of this boson under the global supercharge that defines the elliptic genus in~\eqref{eq:ellgenNS} is a combination of the fermions~$Q^a$. This superpartner is a chiral primary, i.e., it has a zero mode under~$L_0-J_0/2$. 
The derivative in~\eqref{eq:largeN4index} is equivalent to absorbing this fermion zero mode. 
In order to remove the bosonic zero mode, we can pass to the fixed~$\mathrm{U}(1)$-charge ensemble.

We can also understand~\eqref{eq:largeN4index} as an index that counts BPS states of the~$\widetilde A_\gamma$, 
which is the non-linear algebra obtained by quotienting the~$A_\gamma$ algebra by the~$\mathrm{U}(1)$ generator and the four~$Q$ fermionic generators (see~\cref{sec:N4algebra}).
This quotient removes the bosonic as well as the fermionic zero modes from the Hilbert space.
Finally, note that the~$\mathrm{U}(1)$ current and its superpartner under the global supercharge that defines the elliptic genus in~\eqref{eq:ellgenNS} can be mapped to one of the four~$\mathrm{U}(1)$s and its 
superpartner in the~$\asx{T^4}$ theory, upon an Inonu-Wigner contraction. 
The need to absorb the fermion zero mode in that case is also well-known, and plays a role when 
we consider the relation of the index and supersymmetric black holes in that theory~\cite{Ferko:2024uxi}.

The index discussed above is defined in the NS sector. It does not directly capture the black hole microstates, which live in the Ramond sector. 
One might imagine that the right thing to do is to define an index as a trace in the R-sector of the 
large~$\mathcal{N}=4$ algebra, which indeed is feasible~\cite{Gukov:2004fh,Gukov:2004ym}. 
In that case, the R-sector characters of the short representations have a single zero 
and those of the long representations have a double zero at~\hbox{$\rho_+-\rho_-=1$}. 
After absorbing the fermionic zero mode, one obtains the 
R-sector analog of the index~\eqref{eq:largeN4index}. 
We will compute this index in~\cref{sec:nextBTZ}, but there are some subtle points that are worth appreciating at the outset. In particular, BPS states and the above index have unusual properties in the context of the large~$\mathcal{N}=4$ algebra as compared to the smaller two-dimensional superconformal algebras and deserve some comments (much of this material is described in~\cite{Gukov:2004fh}). 
\begin{itemize}[wide,left=0pt]
\item Firstly, since the~$\mathcal{N}=2$ R-current in~\eqref{eq:Neq2subalg} is a~$\gamma$-dependent combination of the two~$\sut$ currents of the  large~$\mathcal{N}=4$ algebra, 
the notion of spectral flow (and hence the notion of NS and R sectors) in the two algebras are different. 
\item Secondly, in the large algebra, the supersymmetric states in the R-sector have positive energies~\eqref{eq:Runitarity}. This can also be seen explicitly using the R-sector characters~\cite{Gukov:2004fh}.   
\end{itemize}

Let us spell out the consequences of these two points. In the~$\mathcal{N}=2$ context, we are used to the fact that chiral primaries in the NS sector map to R-sector supersymmetric ground states under spectral flow.  A general $\mathcal{N}=4$ spectral flow involving a linear combination of the $\sut_\pm$ currents
does not generically map between NS and R sectors. It does so for one unit of spectral flow with one or the other R-symmetry generators. 
For instance, consider  NS sector representations at the unitarity bound with $\sut_\pm$ quantum numbers $\ell_+$ and  $\ell_-$, respectively. 
Their weight $h(\ell_+ ,\ell_-)$ saturates the bound in~\eqref{eq:N4uni}. 
These flow under one unit of $\sut_\pm $ spectral flow (denoted by $\mathscr{S}_\pm$) to the following R-sector representations 
\begin{equation}
\begin{split}
\mathscr{S}_+: \quad \pqty{h(\ell_+, \ell_-), \ell_+, \ell_-}_\mathrm{NS}  
& \; \mapsto \; 
\pqty{h(\ell_+, \ell_-) - \ell_+ + \frac{\mathsf{k_+}}{4}, 
\frac{\mathsf{k_+}}{2} - \ell_+, \ell_- + \frac{1}{2}}_\mathrm{R}  
\\
\mathscr{S}_-: \quad\pqty{h(\ell_+, \ell_-), \ell_+, \ell_-}_\mathrm{NS}  
& \; \mapsto \; 
\pqty{h(\ell_+, \ell_-) - \ell_- + \frac{\mathsf{k_-}}{4},  
\ell_+ + \frac{1}{2}, \frac{\mathsf{k_-}}{2} -\ell_- }_\mathrm{R}  \,.
\end{split}
\end{equation}
Observe therefore that the NS  vacuum representation with $\ell_+ = \ell_- =0$ flows to the R sector 
representation with 
$h_{_\mathrm{R}} = \frac{c}{24} + \frac{1/4}{\mathsf{k}_+ + \mathsf{k}_-}$ (and $(\ell_+,\ell_-) = (\frac12, \frac{\mathsf{k}_-}{2})$ under~$\mathscr{S}_-$).  
Owing to these two features, we should expect an index that is not holomorphic as pointed out in~\cite{Gukov:2004fh}. 
There will be a non-trivial dependence on temperature due to the energy above the threshold. We will therefore find an index that captures the spectrum of short R-sector multiplets in~\cref{sec:nextBTZ}.

\section{The worldsheet string}\label{sec:wsstring}
%
The worldsheet string theory on the geometry~\eqref{eq:dssq} is described by a solvable sigma model. The matter CFT comprises three $\mathcal{N}=1$ WZW models and a free CFT for the $\mathbf{S}^1$ direction. Specifically, we have
\begin{itemize}[wide,left=0pt]
  \item An $\slt$ super-WZW model at level $k = k_+\,\gamma$, which corresponds to the \AdS{3} part of the geometry. This CFT has central charge $c = \frac{9}{2} + \frac{6}{k}$ and currents $J^A$ with $A \in \{\pm,3\}$. The fermions can be decoupled by a field redefinition~\cite{DiVecchia:1984nyg}. This results in a bosonic $\slt$ WZW model with currents $j^A$ at level $k+2$ and three decoupled fermions $\psi^A$ (which themselves form a $\slt$ current algebra with currents $\hat{j}^A$ at level $-2$).
  \item Two $\sut$ super-WZW models at levels $k_\pm$, respectively, corresponding to the two $\mathbf{S}^3$s. They are generated by currents $K^a_\pm$ with $a\in \{\pm,3\}$, and have central charges $c = \frac{9}{2} - \frac{6}{k_\pm}$, respectively. Once again, the field redefinition allows us to decouple the three fermions $\chi^a_\pm$ from each of them, resulting in a bosonic $\sut$ subalgebra (generated by $k^a$) at level $k_\pm-2$ and a fermionic subalgebra (generated by $\hat{k}^a$) at level $+2$.
  \item A free boson and fermion from the $\mathbf{S}^1$, making up the remaining central charge $c = \frac{3}{2}$.
\end{itemize}

Given their solvability, we can put together the torus partition function for each of the aforementioned CFTs and obtain the worldsheet partition function. Before doing so, we should specify the boundary conditions. Let $\{\theta^\pm, \phi_1^\pm, \phi_2^\pm\}$ coordinatize the two $\mathbf{S}^3$s. We will take the Euclidean time circle to be compact with period $\beta$ to work in the thermal \AdS{3} geometry $H_3^+/\mathbb{Z}$. This is already made explicit in our line element~\eqref{eq:dssq}. We will include chemical potentials for angular momentum along the spatial circle in the $H_3^+$, and for the 
R-symmetries for $\mathrm{SU}(2) \times \mathrm{SU}(2)$ rotations along both the three-spheres (but not for the $\mathrm{U}(1)$ isometry along the $\mathbf{S}^1$). Specifically, the coordinates $\{\varphi, \phi_1^\pm,\phi_2^\pm,y\}$ are all periodic with period $2\pi$, and
the twisted boundary conditions around the thermal circle are\footnote{ One could additionally include a $\mathrm{U}(1)$ chemical potential by including a twist along $y$, by requiring the identification also involve $y \to y+ i\,\beta\, \upsilon$. We do not attempt to include this chemical potential. Doing so leads to some confusing features in the worldsheet analysis (similar to issues encountered in obtaining a modular invariant and spacetime supersymmetric partition function for $\asx{K3}$ that was explained in~\cite{Ferko:2024uxi}). }
\begin{equation}
\begin{split}
 & (\tE,\varphi,\phi_1^+, \phi_2^+,\phi_1^-,\phi_2^-,y) \\
 & \; \sim \;
(\tE +\beta,\varphi + i\,\beta\,\mu,\phi_1^+ +i\,\beta\, \nu_1^+, \phi_2^+ + i\,\beta\, \nu_2^+, \phi_1^- +i\,\beta\, \nu_1^-, \phi_2^- + i\,\beta\, \nu_2^-, y )\,.
\end{split}
\end{equation}

It will be useful to introduce modular and elliptic parameters associated with these boundary conditions. We will use
\begin{equation}
\tau \=
\frac{\beta\,\mu + i\,\beta}{2\pi}\,, \qquad q = e^{2\pi i\, \tau}\,,
\end{equation}
to denote the  modular parameter of the spacetime CFT. Furthermore, we set
\begin{equation}
\rho_\pm
\=
\beta\, (\nu_2^\pm - \nu_1^\pm) \,,
\qquad
\overline{\rho}_\pm
\= \beta \, (\overline{\nu}_2^\pm + \overline{\nu}_1^\pm)\,,
\qquad
r_\pm = e^{2\pi i\, \rho_\pm} \,,
\end{equation}
to be the elliptic parameters associated with the $\sut$ chemical potentials.

\subsection{The worldsheet partition function}\label{sec:Zws}

The one-loop partition function of the worldsheet string can be constructed from those of the super-WZW models. We will use
$\ts$ for the worldsheet modular parameter, and define
\begin{equation}
\ts = \ts_1 + i\,\ts_2 \,, \qquad \zs = e^{2\pi i\, \ts}\,,
\end{equation}
reserving $\tau$ and $q$ for the boundary modular parameters, as in~\cite{Ferko:2024uxi}.

We decompose the  $\mathcal{N}=1$ super-current-algebras into bosonic and fermionic pieces as described above. Let us start with the bosonic contribution to the partition function and then discuss the fermionic part, which should be appropriately GSO projected.

\begin{itemize}[wide,left=0pt]
  \item  First, the one-loop partition function of the bosonic $\slt$ WZW was first calculated in~\cite{Gawedzki:1991yu} and was used to obtain the bosonic string partition function in \AdS{3} $\times\, X$ in~\cite{Maldacena:2000kv}. The result at level $k+2$ is given by the modular invariant answer
\begin{equation}\label{eq:Zsl2bos}
\begin{split}
\mathcal{Z}_{\slt}(\ts)
& \=
    \frac{\beta\, \sqrt{k}}{2\pi\, \sqrt{\ts_2}}\,
    \sum_{n,m \in \mathbb{Z}}\,  \mathcal{Z}_{\slt}^{(n,m)}(\ts)  \\
\mathcal{Z}_{\slt}^{(n,m)}(\ts) 
& \=  
    \frac{\exp \left( -\frac{k\,\beta^2}{4\pi\,\ts_2} \, \abs{m- n\,\ts}^2 +
    \frac{2\pi}{\ts_2} \, (\Im(\us_{n,m}))^2\right)}{
    \abs{\vartheta_1(\us_{n,m},\ts)}^2}\,.
\end{split}
\end{equation}
The partition function depends on the holonomies $\us_{n,m}$, which are defined as
\begin{equation}\label{eq:unmdef}
\us_{n,m} = \tau \, (n\,\ts - m )\,.
\end{equation}
In this expression, the factor of $\vartheta_1$ comes from the worldsheet oscillators, while the exponential factor arises from the winding of the strings around the non-contractible thermal circle (while being twisted along the spatial circle). The prefactor originates from the zero mode contribution from the non-compact direction.
\item Next, the contribution of the bosonic $\sut_{k-2}$ WZW model is given by
\begin{equation}\label{eq:Zsusys3}
\mathcal{Z}^{(n,m)}_{\sut}(\ts)
\=
e^{\frac{\pi}{\ts_2}\, (\Im(\rho_{n,m}))^2 \left[k\,\abs{n\,\ts-m}^2 
- (k-2) \right]}  \, \sum_{\ell=0}^{\frac{k}{2}-1} \,
\abs{\frac{\Theta_{2\ell+1}^{(k)}(\rho_{n,m},\ts) -
\Theta_{-2\ell-1}^{(k)}(\rho_{n,m},\ts)}{\vartheta_1
\left( \rho_{n,m}, \ts \right)} }^2\,,
\end{equation}
with
\begin{equation}
\Theta_l^{(k)}(\rho,\tau)
\=
\sum_{n\in \mathbb{Z} + \frac{l}{2k}}\, q^{k\,n^2}\, r^{n \,k} \,,
\qquad
q = e^{2\pi i \tau} \,, \qquad r = e^{2\pi i \rho} \,.
\end{equation}
The $\mathrm{SU}(2)$ chemical potential $\rho$ appears in the answer in the combination
\begin{equation}
\rho_{n,m} \=
\rho \, (n\,\ts -m)\,,
\end{equation}
owing to the twisted boundary conditions, in fact,  exactly as the \AdS{3} chemical potentials appear in the combination~$\us_{n,m}$.
\item
Thirdly, the free boson along the $\mathbf{S}^1$ contributes
\begin{equation}
\mathcal{Z}^{(n,m)}_{\mathrm{u}(1)}(\ts)
\=
\frac{1}{\abs{\eta(\ts)}^2}\, \sum_{p_L,p_R} \, \zs^{p_L^2}\, \zsb^{p_R^2}\,.
\end{equation}
\item Finally, we also have the $bc$ ghost contribution which is
\begin{equation}\label{eq:Zbcghosts}
\mathcal{Z}_{bc}(\ts) \=
\abs{\eta(\ts)}^4\,.
\end{equation}
\end{itemize}
The $\slt$, $\sut$ and free boson partition functions are modular invariant by themselves. The combination of the ghost contribution and the measure factor over the upper-half $\ts$ plane is likewise modular invariant.

Having discussed the contribution from the bosons, we now turn to the fermionic contributions. This depends on the choice of boundary conditions for the fermions around the thermal circle. For periodic boundary conditions for the spacetime fermions, the chiral GSO projection for IIB/A superstring involves the following combination of spin structures
\begin{equation}\label{eq:pfermion}
\begin{split}
\bigl( \mathcal{Z}^\pm_{_\mathrm{PF}} \bigr)^{(n,m)}
 & \=
\Tr_{_\mathrm{NS}}\left(\frac{1-e^{i\pi F}}{2}\, e^{2\pi i \, \ts\, L_0}  \right) -
\Tr_{_\mathrm{R}}\left(\frac{1 \pm e^{i\pi F}}{2}\, e^{2\pi i \, \ts\, L_0}  \right) \\
 & \= \frac12 \Bigl(
\mathcal{Z}^{(n,m)}_3 - \mathcal{Z}^{(n,m)}_4 - \mathcal{Z}^{(n,m)}_2  \mp \mathcal{Z}^{(n,m)}_1 \Bigr) \,.
\end{split}
\end{equation}
The holomorphic characters $\mathcal{Z}_a^{(n,m)}$ represent the trace over the fermion Hilbert space with  different spin structures and are explicitly given by
\begin{equation}\label{eq:Zfermion}
\begin{split}
\mathcal{Z}^{(n,m)}_a
 & \=
\underbrace{\frac{e^{-\frac{\pi}{\ts_2}\, (\Im(\us_{n,m}))^2}\,\vartheta_a(\us_{n,m},\ts)\, \sqrt{\vartheta_a(0,\ts)}}{\eta(\ts)^{\frac{3}{2}}}}_{\slt}
\times
\underbrace{e^{-\frac{\pi}{\ts_2}\, (\Im(\mathfrak{\rho}^+_{n,m}))^2}\frac{\vartheta_a(\rho^+_{n,m},\ts)\, \sqrt{\vartheta_a(0,\ts)}}{\eta(\ts)^{\frac{3}{2}}}}_{\sut_+} \\
 & \qquad  \quad
\times
\underbrace{e^{-\frac{\pi}{\ts_2}\, (\Im(\mathfrak{\rho}^-_{n,m}))^2}\frac{\vartheta_a(\rho^-_{n,m},\ts)\, \sqrt{\vartheta_a(0,\ts)}}{\eta(\ts)^{\frac{3}{2}}}}_{\sut_-} \times
\underbrace{\sqrt{\frac{\vartheta_a(0,\ts)}{\eta(\ts)}}}_{\mathfrak{u}(1)}
\times
\underbrace{\frac{\eta(\ts)}{\vartheta_a(0,\ts)}}_{\beta\gamma} \,.
\end{split}
\end{equation}
Since the trace in the Ramond sector with $(-1)^F$ vanishes due to the presence of fermion zero modes $\mathcal{Z}_1 = 0$, it follows that $\mathcal{Z}_{_\mathrm{PF}}^+ = \mathcal{Z}_{_\mathrm{PF}}^-$. Henceforth, we will drop the distinction between IIB and IIA theories, implicitly focussing on the former for convenience.

With this understanding, the contribution from the left-moving fermions (with periodic boundary condition in spacetime) is given by
\begin{equation}\label{eq:ZperIIB}
\begin{split}
\mathcal{Z}_{_\mathrm{PF}}^{(n,m)}(\ts)
& \=
\frac{e^{-\frac{\pi}{\ts_2} \bqty{(\Im(\mathfrak{u}_{n,m}))^2 \,
+ \,(\Im(\rho^+_{n,m}))^2 \,+\,
(\Im(\rho^-_{n,m}))^2 }}}{\eta(\ts)^4} \\
& \qquad \times
\prod_{\eta_\pm = \pm 1}\;
\vartheta_1\left(\frac{-\mathfrak{u}_{n,m} +\, \eta_+\, \rho^+_{n,m}+
  \eta_-\, \rho^-_{n,m}}{2},\ts  \right) .
\end{split}
\end{equation}
To obtain this result we combined the fermion characters~\eqref{eq:Zfermion} as in~\eqref{eq:pfermion}. The GSO projected combination of characters thus obtained was further simplified using a Riemann identity of theta functions to write the final answer in a Green-Schwarz like form.\footnote{ The choice of IIA GSO projection will lead to the final expression encountered in~\eqref{eq:ZperIIB} with a simple change: $\us_{n,m} \to - \us_{n,m}$. } This expression (projected to $n=0, m=1$ was previously obtained in~\cite{Ashok:2020dnc}. 

Alternately, we can impose antiperiodic boundary conditions for the fermions around the thermal circle.  In this case, the spin-structure sum is performed with a specified set of phases~\cite{Atick:1988si}, which gives the thermal  partition sum
\begin{equation}
\begin{split} \label{eq:AWtwist}
\bigl( \mathcal{Z}^\pm_{_\mathrm{AF}}\bigr)^{(n,m)}
 & \=
\Tr_{_\mathrm{NS}} \left(\left[ \frac{1-e^{i\pi F}}{2} \frac{1+e^{i\pi\,n}}{2} + \, \frac{1+e^{i\pi F}}{2} \, e^{i\pi\,m}\, \frac{1-e^{i\pi\,n}}{2} \right] e^{2\pi i \, \ts\, L_0} \right) \\
 & \qquad
- \Tr_{_\mathrm{R}} \left(\left[ \frac{1\pm e^{i\pi F}}{2} \frac{1-e^{i\pi\,n}}{2} + \, \frac{1+e^{i\pi F}}{2} \, e^{i\pi\,m}\, \frac{1-e^{i\pi\,n}}{2} \right] e^{2\pi i \, \ts\, L_0} \right)\,.
\end{split}
\end{equation}
It is straightforward to write down the result using the characters $\mathcal{Z}_a$ given above. However, as discussed in~\cite{Ferko:2024uxi} the antiperiodic fermion boundary condition can be obtained from the periodic one by tuning the $\mathrm{SU}(2)$ chemical potential (and vice versa). We will therefore focus on the case of the periodic fermions, and simply quote the result for the antiperiodic fermions at the end.

The final result for the worldsheet partition function is then given by
\begin{equation}\label{eq:ZwsIIB}
\begin{split}
\mathcal{Z}_\text{IIB}
 & =
\int_{\mathcal{F}}\, \frac{d^2\ts}{\ts_2}\, \sum_{n,m\in \mathbb{Z}}\,
\mathcal{Z}^{(n,m)}_{\sut}(\ts)\,
\mathcal{Z}^{(n,m)}_{\sut_+}(\ts) \,
\mathcal{Z}^{(n,m)}_{\sut_-}(\ts)\,  \mathcal{Z}_{bc}(\ts)\,
\abs{\mathcal{Z}_{_\mathrm{PF}}^{(n,m)}(\ts)}^2 \,.
\end{split}
\end{equation}
The worldsheet modular integral is over the fundamental domain $\mathcal{F}$ of $\mathrm{PSL}(2,\mathbb{Z})$ in the complex $\ts$-plane. One can check that the integrand is modular invariant once one accounts for the sum over the windings characterized by the 2-tuple $(n,m)$.

\subsection{The single string spectrum}\label{sec:ssspec}

We now proceed to extract the single string spectrum from the worldsheet partition function. In particular, one can exploit the modular invariance of the worldsheet integrand as in~\cite{Zagier:RanSel,Polchinski:1985zf} to convert the double sum over windings $(n,m)$ into a single sum of $m \in \mathbb{Z}_{>0}$, while extending the domain of integration to the strip in the $\ts$ plane.

Implementing this as in~\cite{Maldacena:2000kv} (cf.,~\cite{Ferko:2024uxi} for additional details) we write the worldsheet partition function as
\begin{equation}\label{eq:Zwsfock}
\mathcal{Z}_\text{IIB}(\tau,\rho_+, \rho_-)
\=
-\beta\, \sum_{m=1}^\infty \, f_{_\mathrm{PF}}(m\,\tau,\rho_+,\rho_-) \,,
\end{equation}
The quantity on the r.h.s., $f_{_\mathrm{PF}}(\tau,\rho_+,\rho_-) = f_{_\mathrm{PF}}(\tau,\rho_\pm)$, is the single string free energy. The sum over the free string Fock space in~\eqref{eq:Zwsfock} is the contribution to the spacetime CFT free energy from the thermal $\asx{\stso}$ saddle. Specifically,
\begin{equation}
\mathscr{Z}_{_\mathrm{CFT}}(\tau, \rho_+,\rho_-) \bigg|_\text{thermal \AdS{3}}
\=
\exp(S_\text{tree} + \mathcal{Z}_\text{IIB} + \cdots) \,.
\end{equation}
Here $S_\text{tree}$ is the tree level (genus-$0$) contribution from the worldsheet with no insertions, which contains the Casimir energy of the spacetime CFT.  Thus far, there is no direct computation of this quantity from the worldsheet. We will not attempt to evaluate it here, but assume that it is given by the semiclassical gravity answer when needed.

The single string free energy gives rises to a spectrum with discrete and continuous spacetime CFT weights,
\begin{equation}
f_{_\mathrm{PF}}(\tau,\rho_+,\rho_-)
\=
f_{_\mathrm{PF,disc}}(\tau,\rho_+,\rho_-) + f_{_\mathrm{PF,cont}}(\tau,\rho_+,\rho_-)  \,.
\end{equation}
We will now extract these two pieces from the worldsheet partition function. Since the general notations and ideas are as spelled out in detail in~\cite{Ferko:2024uxi}, we will be brief, and only highlight the salient results.

The main idea is to evaluate the integral over the worldsheet modulus and obtain the single string free energy. To do so, we isolate the contribution from the $\slt$ zero-mode and the classical winding contribution, and write
\begin{equation}\label{eq:fPFintegral}
\begin{split}
f_{_\mathrm{PF}}
 & \=
\frac{\sqrt{k}\,\beta}{2\pi} \, \int_0^\infty\, \frac{d\ts_2}{\ts_2^{\frac{3}{2}}}\, \int_{-\frac{1}{2}}^{\frac{1}{2}}\, d\ts_1\, e^{- \frac{k\,\beta^2}{4\pi\,\ts_2}} \; \mathcal{I}_{_\mathrm{PF}} \,, \\
 & \=
\frac{4}{i\,k} \, \int_{-\infty}^{\infty}\, \zeta\,d\zeta \int_{0}^\infty\, d\ts_2 \, \int_{-\frac{1}{2}}^{\frac{1}{2}}\, d\ts_1\, e^{2i\,\beta\,\zeta - \frac{4\pi}{k} \, \ts_2\,\zeta^2 }\, \mathcal{I}_{_\mathrm{PF}}\,.
\end{split}
\end{equation}
The function appearing in the integrand $\mathcal{I}_{_\mathrm{PF}}$ is the contribution from the worldsheet primaries and oscillators, and is given by
\begin{equation}\label{eq:IPFdefine}
\begin{split}
\mathcal{I}_{_\mathrm{PF}}
 & \=
\abs{\frac{\prod_{\epsilon_\pm = \pm} \, \vartheta_1\left(\frac{\tau \,+\, \epsilon_+\,\rho_+ \,+\, \epsilon_-\, \rho_-}{2},\ts \right)}{\vartheta_1(\tau,\ts)\, \vartheta_1(\rho_+,\ts)\, \vartheta_1(\rho_-,\ts)\, \eta(\ts)^3}}^2 \times \left(\sum_{p_L,p_R}\, \zs^{p_L^2}\, \zsb^{p_R^2}  \right) \\
 & \qquad
\times
\prod_{\epsilon \,\in\, \{\pm\} } \left[
\,  \sum\limits_{\ell_\epsilon=0}^{\frac{1}{2}\,k_\epsilon-1}
\abs{
  \sum_{n_\epsilon\in \mathbb{Z}}\, \zs^{\frac{\left(n_\epsilon\,k_\epsilon+\ell_\epsilon + \frac{1}{2}\right)^2}{k_\epsilon}} \,\left(r_\epsilon^{n\, k_\epsilon + \ell_\epsilon + \frac{1}{2}} -r_\epsilon^{- n_\epsilon\,k_\epsilon - \ell_\epsilon - \frac{1}{2}} \right)}^2
\right] .
\end{split}
\end{equation}
The first line is the oscillator contribution and the $\mathbf{S}^1$ primaries, while the second line is from the primaries, with the second term, which have written using the infinite sum representation, arising from $\sut_{k_\pm-2}$ partition functions.

We can separate out the contribution from the zero modes of the theta functions, and the oscillators and rewrite
\begin{equation}\label{eq:IPFdecompose}
\mathcal{I}_{_\mathrm{PF}}
\=
(q\, \qb)^{-\frac{1}{2}}\, I_0 \times I_{_\mathrm{primaries}} \times I_{_\mathrm{oscillators}} \,.
\end{equation}
The zero mode contribution, which as we shall see contains the spacetime chiral currents, takes the form
\begin{equation}\label{eq:I0def}
\begin{split}
I_0
 & \=
\abs{1+ \frac{\schgS\left(\frac{1}{2},\frac{1}{2},\frac{1}{2}\right) }{1-q}}^2\,.
\end{split}
\end{equation}
We have written the final answer in terms of $\Dta$ characters introduced earlier~\eqref{eq:dtashort}, specifically using the result for low values of spin~\eqref{eq:dtalowspin}.
The contributions from the  $\mathfrak{u}(1)$ and  $\sut$ primaries (with the latter assembled into $\sut$ characters) can be checked to be
\begin{equation}\label{eq:Iprdef}
I_{_\mathrm{primaries}}
\=
\left(\sum_{p_L,p_R}\, \zs^{p_L^2}\, \zsb^{p_R^2}  \right)\, \times
\prod_{\epsilon \,\in\, \{\pm\} }
\left(\sum_{\ell_\epsilon=0}^{\frac{k_\epsilon}{2}-1} \abs{\sum_{n_\epsilon\in\mathbb{Z}} \zs^\frac{\left(n_\epsilon\,k_\epsilon +\, \ell_\epsilon + \frac{1}{2}\right)^2}{k_\epsilon} \, \chi_{n_\epsilon\, k_\epsilon + \ell_\epsilon}(r_\epsilon)}^2 \right).
\end{equation}
Finally, the oscillator contribution takes the form
\begin{equation}\label{eq:Iosc}
I_{_\mathrm{oscillators}}
\=
\abs{\prod_{n=1}^\infty \,
  \dfrac{\prod\limits_{\epsilon_\pm \,= \,\pm 1}
    \left(1- \left(q\,r_+^{\epsilon_+}\, r_-^{\epsilon_-}\right)^\frac{1}{2}\, \zs^n \right)  \left(1- \left(q\,r_+^{\epsilon_+}\, r_-^{\epsilon_-}\right)^{-\frac{1}{2}}\, \zs^n\right) }{(1-\zs^n)^2\, \prod\limits_{\epsilon=\pm 1} \, \left(1-q^\epsilon\,\zs^n\right)\, \left(1-r_+^\epsilon\,\zs^n\right) \,   \left(1-r_-^\epsilon\,\zs^n\right)  }}^2 .
\end{equation}

The  worldsheet modular integral can be evaluated using the strategy explained in~\cite{Maldacena:2000kv}. The issue is the presence of poles in the region of integration from the $\slt$ oscillators, the factor $(1-q^{-1}\, \zs^n)$ in the denominator of $I_{_\mathrm{oscillators}}$. As elaborated in~\cite{Ferko:2024uxi} (see Appendices therein), there is a useful expansion of $\eta(\ts)^3/\vartheta_1(\tau,\ts)$ that can be exploited for our purposes, cf.,~\eqref{eq:theta1exp} (see also~\cite{Ashok:2020dnc}). We therefore separate out the contribution from the $\slt$ and $bc$ ghost part, and expand\footnote{ We have written this expansion in this fashion to facilitate comparison with the original analysis of~\cite{Maldacena:2000kv}, since $N$ is akin to the total number of oscillator excitations on the worldsheet. A more natural expansion of the infinite product is given in~\eqref{eq:theta1exp}. The two expressions differ by a factor of the Dedekind eta function. The powers are related by $ N  = \frac{1}{2}\pqty{ r\,(r+1) - n(n+1) + w (w+1)} +m$ (where $\zs^m$ arises from the $\eta(\ts)$), using which the coefficient series $\mathbf{P}_N^w(q)$ can be deduced. } 
\begin{equation}
\prod_{n=1}^{\infty}\, \frac{1-\zs^n}{(1-q\,\zs^n)\, (1-q^{-1}\, \zs^n)}
\=
\sum_{w=0}^\infty\, q^w\, \zs^{-\frac{1}{2}\, w\,(w+1)}\, \sum_{N\in \mathbb{Z}}\, \mathbf{P}_{N}^w(q)\, \zs^N\,.
\end{equation}
The remaining factors in $I_{_\mathrm{oscillators}}$ can be Fourier expanded with positive powers of $\zs$, with 
coefficients which are $\sut$ characters $\chi_{j_\pm}(r_\pm)$. Combining the two expansions, we arrive at
an expression of the following form
\begin{equation}\label{eq:IoscP}
I_{_\mathrm{oscillators}}
\=
\abs{\sum_{w=0}^\infty \,\sum_{N\in \mathbb{Z}}\, \mathbf{Q}_{N}^w  (q,r_+,r_-)\, q^w  \, \zs^{-\frac{1}{2}\, w (w+1) + N} }^2\,.
\end{equation}
The parameter $w\in \mathbb{Z}_{\geq 0}$ labels the spectral flow sectors, with $w=0$ being the unflowed sector, and $N \in \mathbb{Z}$ is the net level of the oscillators.

Given this decomposition, we can write the integrand~\eqref{eq:fPFintegral} as
\begin{equation}
\begin{split}
\mathcal{I}_{_\mathrm{PF}}
 & \=
(q\,\qb)^{-\frac{1}{2}}\, I_0 \;
\widehat{\sum} \,
(q\, \qb)^w \,  \mathsf{Q}_{N,\overline{N}}^w\,
\times 	 (\zs\, \zsb)^{-\frac{1}{2}\, w (w+1) } \,
\zs^{p_L^2 + h_{\sut} + N} \; \zsb^{p_R^2 + \overline{h}_{\sut} + \overline{N}} \,.
\end{split}
\end{equation}
To write this expression succinctly we  have introduced
\begin{itemize}[wide,left=0pt]
  \item  The weights for the net  contribution $\sut$ primaries:
        \begin{equation}
        \begin{split}
        h_{\sut}
         & \=
        \frac{(n_+\,k_+ +\ell_+ + \frac{1}{2})^2}{k_+} + \frac{(n_-\,k_- + \ell_- + \frac{1}{2})^2}{k_-} \,, \\
        \overline{h}_{\sut}
         & \=
        \frac{(\overline{n}_+\, k_+ +\ell_+ + \frac{1}{2})^2}{k_+} + \frac{(\overline{n}_-\, k_- +\ell_- + \frac{1}{2})^2}{k_-} .
        \end{split}
        \end{equation}
  \item A compact notation for the product of the left and right moving oscillator expansions
        \begin{equation}\label{eq:QNNbwdef}
        \mathsf{Q}_{N,\overline{N}}^w
        \; \equiv \;
        \mathbf{Q}_{N}^w  (q,r_+,r_-)\,
        \overline{\mathbf{Q}}_{_{\overline{N}}}^w  (\qb,\overline{r}_+,\overline{r}_-)
        \left[\prod\limits_{\epsilon = \pm 1} \; \chi_{n_\epsilon\,k_\epsilon + \ell_\epsilon} (r_\epsilon) \, \chi_{\overline{n}_\epsilon k_\epsilon + \ell_\epsilon}(\overline{r}_\epsilon) \right] \,;
        \end{equation}
  \item A compact  notation to collectively denote summation over the various quantum numbers
        \begin{equation}
        \widehat{\sum} \; \equiv \;
        \sum\limits_{w=0}^\infty\
        \sum\limits_{N,\overline{N}\in \mathbb{Z}}\
        \sum\limits_{n_\pm,\overline{n}_\pm \in \mathbb{Z}} \
        \sum\limits_{p_L,p_R}  \
        \sum\limits_{\ell_+\,=\, 0}^{\frac{k_+}{2}-1} \  \sum\limits_{\ell_-\, =\, 0}^{\frac{k_-}{2}-1}   \;.
        \end{equation}
\end{itemize}

The integral over $\ts$ can now be done straightforwardly. The integral over $\ts_1$ imposes level matching demanding that
\begin{equation}\label{eq:levelmatch}
p_L^2 + h_{\sut} + N
\=
p_R^2 + \overline{h}_{\sut} + \overline{N} \,,
\end{equation}
while the integral over $\ts_2$ is elementary and results in
\begin{equation}\label{eq:susyfsp}
\begin{split}
f_{_\mathrm{PF}}
 & \=
\frac{k}{4\pi}\, I_0\,
\widehat{\sum} \, \delta_\text{LM}\;  (q\, \qb)^{w-\frac{1}{2}} \,
\mathsf{Q}_{N,\overline{N}}^w
\times \int_{-\infty}^{\infty}\, \frac{d\zeta}{i\pi} \, \frac{\zeta\,e^{2i\,\beta\,\zeta}}{\zeta^2 + \Delta_w^2} \Bigl( e^{-\frac{2\beta}{w+1}\,\frac{\zeta^2  + \Delta_w^2}{k}} - e^{-\frac{2\beta}{w}\, \frac{\zeta^2 + \Delta_w^2}{k}} \Bigr) \,.
\end{split}
\end{equation}
In this expression $\delta_\text{LM}$ is an insertion of the level matching constraint as a delta function~\eqref{eq:levelmatch}, as well as
\begin{equation}\label{eq:DeltaSS}
\Delta_{w}^2
\=
k\left[p_L^2 + h_{\sut} + N - \frac{1}{2}w(w+1)\right] \,.
\end{equation}

The final remaining integral over the  auxiliary parameter $\zeta$ captures both the discrete and continuous parts of the spectrum. To extract them, we deform the contour away from the real-axis into the upper-half $\zeta$  plane. We shift by an amount that depends on the spectral flow level; $\zeta \to s + \frac{i}{2}\, k\, (w+1)$. In doing so, we pick up residues from the poles at
\begin{equation}
\zeta_* =  i\,\Delta_w \,.
\end{equation}
These will correspond to the discrete states. At a given spectral flow level, a restricted number of poles contribute; see~\eqref{eq:Deltarange} below for the specific bounds. The  remaining integral along $s \in \mathbb{R}$ captures the continuum states. The parameter $s$ itself is akin to a spectral parameter and corresponds to the radial momentum of states in the \AdS{3} geometry.

We can now state the final result for the discrete and continuous parts of the spectrum. The single string free energy comprises
\begin{equation}\label{eq:fdSS}
\begin{split}
f_{_\mathrm{PF,disc}}(\tau, \rho_\pm)
 & \=
I_0 \, \widehat{\sum} \, \delta_\text{LM}\,
\mathsf{Q}_{N,\overline{N}}^w \; (q\, \qb)^{w-\frac{1}{2} + \Delta_w} \,.
\end{split}
\end{equation}
The contribution of the discrete states in the $w^\mathrm{th}$ spectral flowed sector is subject to the constraint\footnote{We have written the unitary bound proposed in~\cite{Maldacena:2000hw} as strict inequalities. 
For generic $k$ one can relax this, since $\Delta_w$ does not saturate the upper or lower bounds. 
However, since the situation will be different for $k=1$, we have written a clean version that can be applied to any $k$.} 
\begin{equation}\label{eq:Deltarange}
\frac{k}{2}\, w  \; < \; \Delta_{w}
\; < \; \frac{k}{2}\, (w +1) \,.
\end{equation}
We have non-vanishing contribution from only a subset of poles in a fixed spectral flow sector because the 
residues are subject to cancellations between the two terms in the parenthesis in~\eqref{eq:susyfsp}. Moreover,~\eqref{eq:Deltarange} also ensures that the spacetime CFT spectrum satisfies unitarity constraints.
The continuum spectrum is determined in terms of a  density of states $\bm{\varrho}(s; q,\qb,r_\pm,\overline{r}_\pm)$, and takes the 
form\footnote{We have written this slightly differently from the expression presented in~\cite{Ferko:2024uxi}, where certain factors were additionally absorbed into the spectral density. }
\begin{equation}\label{eq:fPFcont}
f_{_\mathrm{PF,cont}}(\tau, \rho_\pm)
\=
I_0\,\sum_{w=1}^\infty\, \int_{-\infty}^\infty\,\frac{ds}{i\pi}\, \bm{\varrho}(s; q,\qb,r_\pm,\overline{r}_\pm)\, (q\,\qb)^{\frac{k\,w}{4} + \frac{1}{w\,k}\, \left(s^2 + \Delta_w^2\right) + w - \frac{1}{2}}\,.
\end{equation}
The precise form of this density can be determined as in~\cite{Maldacena:2000kv}. As we won't need it for the rest of our discussion, we don't therefore work out its precise form.

\subsection{Spacetime currents \& supergravity states}\label{sec:specCFT}

Now that we have obtained the worldsheet partition function, we have a handle on
the spectrum of excitations about the \AdS{3} geometry. As we have indicated in our preceding discussion, this information should be consistent with the canonical quantization of the worldsheet theory. Specifically, we should see that the superisometries get enhanced to the appropriate asymptotic symmetry algebra, and furthermore, we recover the spectrum of states seen in supergravity.

Let us focus on the discrete part of the spectrum, ignoring oscillator excitations $N=0$, and work in the $w=0$ spectral flow sector. From~\eqref{eq:fdSS} we note that the conformal dimension of such states is
\begin{equation}
\Delta_{w=0} - \frac{1}{2} = \sqrt{k\, p_L^2 + k \left( \frac{(n_+\,k_+ +\ell_+ + \frac{1}{2})^2}{k_+} + \frac{(n_-\,k_- + \ell_- + \frac{1}{2})^2}{k_-} \right)} -\frac{1}{2} \,,
\end{equation}
and we are required by~\eqref{eq:Deltarange} to ensure that
$0 < \Delta_{0} < \frac{k}{2}$. Let us further simplify to states with
$p_L = p_R = 0$, and also switch off the spectral flow for the $\sut$ states setting $n_\pm =0$. Doing so, we trivially satisfy the level matching condition, and the states are labeled by the two $\sut$ spins $\ell_\pm$. Their spacetime conformal weight is
\begin{equation}
\hs = \sqrt{\frac{k}{k_+} \, \left(\ell_+ + \frac{1}{2} \right)^2 + \frac{k}{k_-}\, \left( \ell_- + \frac{1}{2}\right)^2} -\frac{1}{2} \,.
\end{equation}

We claim that this class of states precisely reproduces the supergravity spectrum obtained in~\cite{Eberhardt:2017fsi}. One can verify this as follows. Consider their contribution to the  single string free energy, which is
\begin{equation}
\begin{split}
f_{_\mathrm{PF,disc}}(\tau, \rho_\pm) \bigg|_\text{sugra}
 & \=
I_0 \,  \sum\limits_{\ell_+= 0}^{\frac{k_+}{2} -1}\,
\sum\limits_{\ell_- =0}^{\frac{k_-}{2} -1}\,
(q\,\qb)^{\hs}\, \abs{ \chi_{\ell_+}(r_+)\, \chi_{\ell_-}(r_-)}^2
\end{split}
\end{equation}
This expression can be recast into a more familiar form using~\eqref{eq:I0def} and the easily verifiable identity of $\Dta$ characters,
\begin{equation}
\schgS\left(\frac{1}{2},\frac{1}{2},\frac{1}{2}\right) \, q^h\, \chi_{\ell_+}(r_+)\, \chi_{\ell_-}(r_-)
\=
\schgL(h,\ell_+,\ell_-) - \, q^h \, (1-q) \, \chi_{\ell_+}(r_+)\,\chi_{\ell_-}(r_-)  \,.
\end{equation}
Implementing this, we arrive at
\begin{equation}
\begin{split}
f_{_\mathrm{PF,disc}}(\tau, \rho_\pm) \bigg|_\text{sugra}
 & \=
\sum\limits_{\ell_+= 0}^{\frac{k_+}{2} -1}\,
\sum\limits_{\ell_- =0}^{\frac{k_-}{2} -1}
\abs{\schgL(\hs,\ell_+,\ell_-)}^2 \,.
\end{split}
\end{equation}
As promised, this is indeed the expected supergravity spectrum~\eqref{eq:sugraalt}.  As discussed in~\cref{sec:sugraspectrum} the diagonal elements of this sum capture the BPS states, while the off-diagonal elements are long multiplets, which nevertheless exist in supergravity. The $\sut$ spins are bounded from above here owing to the stringy exclusion principle, a feature that would not be visible working just in the low energy effective description. All told, we reproduce the spectrum obtained in~\cite{Eberhardt:2017fsi}.

Now that we have recovered the supergravity spectrum, let us examine the boundary currents. While a generic state has non-vanishing twist, $\hs - \hsb \neq 0$, we see that the states with $\ell_+  = \ell_- =0 $ have a very simple contribution given by the zero modes of the worldsheet oscillators, and contribute
\begin{equation}
I_0 \=
1 + \frac{\schgS\left(\frac{1}{2},\frac{1}{2},\frac{1}{2}\right) }{1-q} +
\frac{\schgbS\left(\frac{1}{2},\frac{1}{2},\frac{1}{2}\right) }{1-\qb}
+ \frac{\schgS\left(\frac{1}{2},\frac{1}{2},\frac{1}{2}\right) \, \schgbS\left(\frac{1}{2},\frac{1}{2},\frac{1}{2}\right)}{\abs{1-q}^2}\,.
\end{equation}
The first term is the contribution of the spacetime CFT vacuum state. The holomorphic and antiholomorphic piece encode the left and right moving currents, respectively. The final term corresponds to the spacetime current-current bilinears as discussed in~\cite{Ferko:2024uxi}. Recalling~\eqref{eq:dtalowspin} we recognize that the currents are precisely what we expect: the solitary state of holomorphic weight $2$ which is in $\sut$ singlets is the boundary graviton, while the $\sut$ currents are the states with weight $1$ transforming in the adjoint. The states with weight $\frac{3}{2}$ are the supercurrents.  The singlet current with contribution $q$ and states of weight $\frac{1}{2}$ are the contribution from the $\mathrm{U}(1)$ current associate with the $\mathbf{S}^1$ isometry, and its fermion partners. This multiplet spans out the content of the superisometries of the geometry.

Focusing on the holomorphic contribution, we can deduce the contribution of the currents to the multi-string partition function. This is given by
\begin{equation}\label{eq:stringvacchar}
\exp \biggl(\, \sum_{j=1}^\infty\, \frac{1}{j} \, I_{0,\text{hol}}(j\,\tau, j\,\rho_\pm) \biggr)
 \=
\dfrac{\vartheta_4^2\left(\frac{\rho_+ + \rho_-}{2},\tau\right)\, \vartheta_4^2\left(\frac{\rho_+ - \rho_-}{2},\tau\right)}{\eta(\tau)^6\, \vartheta_1(\rho_+,\tau)\,\vartheta_1(\rho_-,\tau)} \,
\dfrac{(r_+^{\frac{1}{2}} -r_+^{-\frac{1}{2}})\, (r_-^{\frac{1}{2}} -r_-^{-\frac{1}{2}})\, (1-q)}{ \prod\limits_{\epsilon_\pm = \pm 1} \left(1 - r_+^{\frac{1}{2}\,\epsilon_+}\, r_-^{\frac{1}{2}\,\epsilon_-} q^{\frac{1}{2}}\right)  } 
\end{equation}
In other words, we recover the vacuum character of the large $\mathcal{N} =4$ superconformal algebra, albeit just the part where the sum over the integer parameters $m,n$  is truncated down to $n=m =0$, i.e., 
\begin{equation}
\exp \biggl(\, \sum_{j=1}^\infty\, \frac{1}{j} \, I_{0,\text{hol}}(j\,\tau, j\,\rho_\pm) \biggr)
 \=
\schr{\text{NS}}^\gamma(q,r_+, r_-,u) \bigg|_{n=m=0} \,. 
\end{equation}
Specifically, the function $\mu_{_\mathrm{NS}}(q,r_+,r_-)$ defined in~\eqref{eq:mudef}, for short multiplets with $n=m =0$ is simply
\begin{equation}
\mu_{_\mathrm{NS}}(q,r_+,r_-) \bigg|_{n=m=0}
\=
\sum_{\epsilon_\pm = \pm 1} \, \frac{\epsilon_+\, \epsilon_-}{1- r_+^{\frac{1}{2}\, \epsilon_+}\, r_-^{\frac{1}{2} \, \epsilon_-}\, q^{\frac{1}{2}}}
\=
\dfrac{(r_+^{\frac{1}{2}} -r_+^{-\frac{1}{2}})\, (r_-^{\frac{1}{2}} -r_-^{-\frac{1}{2}})\, (1-q)}{ \prod\limits_{\epsilon_\pm = \pm 1} \left(1 - r_+^{\frac{1}{2}\,\epsilon_+}\, r_-^{\frac{1}{2}\,\epsilon_-} q^{\frac{1}{2}}\right)  }\,.
\end{equation}

While the thermal \AdS{3} geometry only reproduces a piece of the full superconformal vacuum character, one can recover the full character by summing over a set of saddle geometries. As noted at the outset,  the two $\mathbf{S}^3$s are fibered over the \AdS{3} base because of the twisted boundary conditions associated with the $R$-charge chemical potentials. In the effective three-dimensional gravitational dynamics, the fibration implies that there are background values for $\sut$ Chern-Simons gauge fields $A$. As explained in~\cite{Kraus:2006nb} the chemical potential fixes the antiholomorphic component of the gauge field along the \AdS{3} boundary, but the holomorphic part is only fixed up to an additive integer. These are precisely the integers the superconformal character sums over. In CFT parlance, they correspond to spectral flow in the two $\sut$ subalgebras. For the problem at hand, the spectral flow amounts to an average over both the elliptic parameters $\rho_\pm$,
\begin{equation}\label{eq:sflowsum}
f(r_+, r_-)
\; \mapsto \;
\sum_{n,m\in \mathbb{Z}}\, q^{k_+\, n^2 + k_-\, m^2 }\, r_+^{k_+\, n}\, r_-^{k_-\,m}\, f(q^{2\,n}\, r_+, q^{2\,m}\, r_-)\,.
\end{equation}
Upon implementing this, we see that the full large $\mathcal{N} = 4$ vacuum character is recovered.

To wrap up this discussion, let us finally note that states with $p_L \neq 0$ or with $\sut$ spectral flow index $n_\pm \neq 0$ have spacetime weights of $\order{\sqrt{k}}$. They therefore lie outside the supergravity regime which is attained for $k \gg 1$.

\subsection{The regime of string scale geometry}\label{sec:stringk1}

The results we have derived hold for generic values of $k_\pm$. In the RNS formalism, the minimum value of these parameters that we can access is $k_\pm =2$. This bound arises because of the shift in the level of the bosonic current algebra when we decouple the fermions from the $\mathcal{N}=1$ supercurrent algebra. When $k_\pm =2$ the $\sut_{k_\pm}$ partition function~\eqref{eq:Zsusys3} simplifies considerably; $\mathcal{Z}_{\sut} =1$, with only the identity representation surviving. The contribution from the fermions is unchanged since they were decoupled in the first place.

First, consider the case where $k_- =2$ whence $k = \frac{2\,k_+}{2+ k_+} $, and let $k_+$ be generic.
(Of course one can interchange the two roles using the $k_+ \leftrightarrow k_-$ symmetry.)
This forces $\ell_- =0$, and therefore the BPS multiplets that survive have $\ell_- = \ell_+ =0$. 
Thus, only the chiral currents which correspond to the boundary supergraviton multiplet lie in the short multiplets. 
However, non-BPS multiplets $(\ell_+ \geq \frac{1}{2}, \ell_- = 0)$ contribute to the discrete spectrum of the spacetime CFT.

A more interesting limit is $k_+ = k_- =2$, whence $k=1$. Now both the $\mathbf{S}^3$s only contribute states from the identity representation to the bosonic part. Replacing the bosonic $\sut$ partition functions by unity,  the single string contribution to the spacetime free energy therefore simplifies to
\begin{equation}\label{eq:fPFintegralk1}
\begin{split}
f_{_\text{PF}}
 & \=
-4\,i \, \int_{-\infty}^{\infty}\, \zeta\,d\zeta \int_{0}^\infty\, d\ts_2 \, \int_{-\frac{1}{2}}^{\frac{1}{2}}\, d\ts_1\, e^{2i\,\beta\,\zeta - 4\pi \, \ts_2\,\zeta^2 }\, \mathcal{I}_{_\text{PF}}\,, \\
\mathcal{I}_{_\text{PF}}
 & \=
\abs{\frac{\prod\limits_{\epsilon_\pm = \pm} \, \vartheta_1\left(\frac{\tau \,+\, \epsilon_+\,\rho_+ \,+\, \epsilon_-\, \rho_-}{2},\ts \right)}{\eta(\ts)^6}
\, \times \, \frac{\eta(\ts)^3}{\vartheta_1(\tau,\ts)}}^2 \times \left(\sum_{p_L,p_R}\, \zs^{p_L^2}\, \zsb^{p_R^2}  \right) .
\end{split}
\end{equation}
The oscillator pieces have been rewritten slightly to facilitate expansion of the theta functions. Note that there is now a zero-point energy  of $\frac{1}{4}$ from this oscillator part. This is similar to the vacuum contribution for general $k$, where this factor comes from the sum over spins.

We can now expand the integrand and complete the integral over the worldsheet modular parameter $\ts$ as before, and obtain  
\begin{equation}\label{eq:k1fpf}
\begin{split}
f_{_\mathrm{PF}}(\tau, \rho_+,\rho_-)
 & \=
I_0 \, \sum\limits_{N,\overline{N} \,\in\, \mathbb{Z}}\;
\sum\limits_{p_L,p_R}\;\sum\limits_{w=0}^\infty\,
\delta_{N + p_L^2 - p_R^2 , \overline{N}}\ 
\widetilde{\mathsf{Q}}_{N,\overline{N}}^w  \;
(q\, \qb)^{w -\frac{1}{2}} \, \mathfrak{J}_{k=1} \,,\\
\mathfrak{J}_{k=1} & \= 
\int_{-\infty}^\infty\, d\zeta\, \frac{\zeta\;e^{2i\beta\,\zeta}}{\zeta^2
+ \Delta_w^2} \bqty{
  (q\,\qb)^{\frac{\zeta^2+ \Delta_w^2}{w+1}} -  (q\,\qb)^{\frac{\zeta^2+ \Delta_w^2}{w}}
}\,.
\end{split}
\end{equation}
In this expression~$\Delta_w$ is given by setting $h_{\sut} = \frac{1}{4}$ and
$k=1$ in~\eqref{eq:DeltaSS}, i.e.,
\begin{equation}\label{eq:DeltaSS1}
\Delta_w^2 \= p_L^2  + N + \frac{1}{4} - \frac{1}{2}\, w \,(w+1)\,.
\end{equation}
Furthermore, $\widetilde{\mathsf{Q}}$ indicates the contributions of the
oscillators, similar to $\mathsf{Q}$ in~\eqref{eq:QNNbwdef}, except
that now the $\sut$ oscillators are missing. The derivation of this result is explained in~\cref{sec:tensionless}.\footnote{We caution the reader that in the parameterization used here certain features are not completely transparent. In particular,  the bounds on spacetime dimensions naively appear to be negative since a priori $N \in \mathbb{Z}$. Showing that this does not in fact transpire requires one to expand the $\slt$ partition function with care. We explain how this can be done  and demonstrate spacetime unitarity~\cref{sec:tensionless}, which we encourage the interested reader to consult.\label{fn:caveatB} } 
 
We should complete the integral to extract the spacetime CFT spectrum.  Proceeding as before, we shift the integration contour for $\zeta$ into the positive half plane, defining $\zeta \to s + \frac{i}{2}\,(w+1)$, or 
$\zeta \to s + \frac{i}{2}\, w$. Specifically, for $w\neq 0$ we shift the contour for the two terms in $\mathfrak{J}_{k=1}$ by different amounts, with only poles in the interim range contributing. Doing so, 
one picks up the poles from the zeros of $\zeta^2 + \Delta_w^2$ and leaving behind a continuum piece. The poles will contribute to the discrete states provided we satisfy an analog of the constraint~\eqref{eq:Deltarange}, which in the present case demands
\begin{equation}\label{eq:k1constraint}
\frac{w}{2}\; < \; \sqrt{p_L^2 + N + \frac{1}{4} -\frac{1}{2}\, w \, (w+1) } \;<\; \frac{w+1}{2} \,.
\end{equation}

We re-express~\eqref{eq:k1constraint} as a constraint on allowed values of $p_L^2 + N$ in different spectral flow sectors as follows,
\begin{equation}
\frac{(3\,w-1)\,(w+1)}{4} \,<\,  p_L^2 + N \,<\, \frac{w\,(3\,w+4)}{4}\,,
\end{equation}
for a pole in the $\zeta$-plane with $\frac{w}{2} \leq \Im(\zeta) \leq \frac{w+1}{2}$ to contribute. 
When $w=0$, we have no solutions to~\eqref{eq:k1constraint}. The choice $p_L = N = 0$  is excluded by the strict inequality, though it would have saturated the upper bound.   
As we shall see below, the pole will lie on the shifted contour of integration, signalling the merger with the continuum modes. 
We find this curious, since not only is there no spectral gap, but the continuum also extends all the way down to the identity operator.  
Likewise, for $w=1$, there is no discrete contribution, since the only potential choice $N=1$ and $p_L=0$, again, lies at the edge of admissibility. 
However, for higher values of $w$ there are discrete states,  e.g., $w=2$ admits $N=4$, while $w=3$ admits $N=9$, etc. 
In fact, with increasing values of $w$ we potentially find more solutions as the difference between the upper and lower bound scales linearly with~$w$. 

Let us next examine the continuum part of the spectrum and write out their contribution to the single string free energy.
As before, for a given~$w$, we shift the contour as $\zeta =s + i\,\frac{w+1}{2}$. 
Upon collecting terms with similar scaling weights, we 
find the $k=1$ analog of~\eqref{eq:fPFcont}.
The explicit expression for the continuum spectrum, from which we can read off the density of states~$\bm{\varrho}(s; q,\qb,r_\pm,\overline{r}_\pm)$, is 
%
\begin{equation}\label{eq:fpcontk1}
\begin{split}
f_{_\mathrm{PF,cont}}
& 
    \,\supset \,  I_0\,	\sum_{w=1}^\infty \,\sum_{N, \,p_L,\,p_R}\; \delta_\mathrm{LM}\;
    \int\limits_{-\infty}^\infty\, \frac{ds}{i\pi} \; (q\,\qb)^{E_w(s) } \,  \left[ 
    \frac{ \left(s+\frac{i\,w}{2}\right) \widetilde{ \mathsf{Q}}^{w-1}_N}{\left(s+\frac{i\,w}{2}\right)^2 + \Delta^2_{w-1} } - 
    \frac{\left(s+\frac{i\,w}{2}\right) \widetilde{\mathsf{Q}}^w_N}{ \left(s+\frac{i\,w}{2}\right)^2 +\Delta_w^2 } 
    \right] \,,
\end{split}
\end{equation}
where we have defined 
\begin{equation}
E_w(s) 
\=  
\frac{3}{4}\, w -1 + \frac{1}{w}  \left(s^2+ \frac{1}{4} + N+ p_L^2\right) \,. 
\end{equation}
The spacetime CFT conformal dimensions $h_{_\mathrm{CFT}}$ and $\overline{h}_{_\mathrm{CFT}}$ are determined by $E_w(s)$ and the powers of $q$, $\qb$ arising from the coefficient functions $\widetilde{\mathsf{Q}}^w_N$.

We combined terms from the $(w-1)^\mathrm{st}$ and $w^\mathrm{th}$ spectral flowed sectors to pick out those with weight $E_w(s)$. Note also that the range for the parameter $s$ labeling such representations is $s\in \mathbb{R}$. In comparing this to the expressions in the literature (e.g., the continuum energies quoted in~\cite{Gaberdiel:2018rqv,Gaberdiel:2024dva}) one should be aware that $N\in \mathbb{Z}$ here refers to the net excitation of the fermionic and $\slt$ determinant factors. These determinants have a Laurent expansion owing to the presence of singularities from the $\slt$ part of the characters. Furthermore, our choice of $N$ can be shifted, for example, by absorbing part of the $w$ dependence into $N$. On the other hand, much of the literature uses the canonical quantization of the string, and uses $N$ to refer to the net collection of $\slt$ creation operators (thus a positive level). A useful sanity check is that for $w=1$ and no oscillator excitations, our result agrees with that obtained in the literature.

There are again some curious features of this answer: the integrand in~\eqref{eq:fpcontk1}, which scales as $(q\,\qb)^{E_w(s)}$ contributes to both the left and right-moving weights. So none of these states will be chiral unless $E_w(s) = 0$. The only case this can happen is for $w=1$ along with $p_L = N =0$, and that too only at $s=0$. Then the chiral contribution in $I_0$ will pick out the pieces we identified earlier as the left and right moving boundary gravitons. But this is precisely the missing discrete series representation noted above, albeit buried under the continuum. Moreover, the chiral part of $I_0$ has maximum twist $2$. 

As noted in~\cref{fn:caveatB} the above argument is a bit quick, since we have not carefully analyzed the spacetime weights arising in the spectrum. A more precise argument using a clean expansion of the $\slt$ partition function is presented in~\cref{sec:tensionless}. As we outline there, the constraints from the spacetime unitarity bound~\eqref{eq:Deltarange} prove sufficiently constraining. Specifically, in the parameterization used here it prevents too negative values of $N$ and $\overline{N}$ from appearing in the spectrum.

The spectrum of the tensionless string at $k=1$ described in~\cite{Gaberdiel:2024dva} (building upon results of~\cite{Gaberdiel:2018rqv}) does not involve any discrete state representations of $\slt$ (or their spectral flows).
In addition, one expects there to be a higher spin symmetry, which should have been manifested by the presence of higher spin characters. Our expression for the partition function, in fact, points towards the absence of higher spin symmetries in this limit, and also indicates the presence of certain discrete states.

We believe the resolution to the difference between the spectrum obtained herein by taking the $k\to1$ of the RNS partition function~\eqref{eq:ZwsIIB}, and that proposed in~\cite{Gaberdiel:2024dva} to lie in the fact that there are two different CFTs in play. One of them is the theory we consider obtained by simply setting $k=1$ in our general expressions. The other ought to be a distinct CFT with no discrete series representations, and only continuum states. As indicated in~\cref{sec:intro} there is  evidence for this claim. Given this, one ought to better understand the theory described herein.

\section{Near-extremal BTZ black holes}\label{sec:nextBTZ}

We now have the one-loop partition function for strings propagating on
$\mathbb{H}^3/\mathbb{Z}$ at hand. While we have thus far interpreted it as
computing the spectrum of strings on the Euclidean thermal \AdS{3} background,
we can equivalently interpret the result as providing us insight into the
spectrum around the Euclidean BTZ geometry. The two geometries share the same
bulk manifold~\eqref{eq:dssq}, but the boundary circles $\tE$ and
$\varphi$ are swapped. From the spacetime CFT perspective, this is an S-modular
transform $\tau \to -\frac{1}{\tau}$.

To interpret our result in terms of the
BTZ spectrum, we should, in addition, account for the spacetime fermion boundary
conditions. We recall that the choice of periodic or antiperiodic fermions is
equivalent to computing the spacetime CFT Hilbert space trace, with or
without the insertion of $(-1)^{F_s}$, with $F_s$ being the spacetime fermion
number. Furthermore, since the thermal \AdS{3} vacuum corresponds to the NS sector
ground state, under the S-modular transform we encounter the following BTZ partition
functions:
\begin{subequations}\label{eq:btzNSRsusy}
  \begin{align}
  \mathscr{Z}^\mathrm{NS}_{_\mathrm{CFT}}(\tau, \rho_{+}, \rho_-, \overline{\tau},\overline{\rho}_+ , \overline{\rho}_-) \Big{|}_\mathrm{BTZ}
   & \=
  \Tr_{_\mathrm{NS}}
  \pqty{q^{L_0}\, r_+^{J_0^{3,+}}\,  r_-^{J_0^{3,-}}\, \overline{q}^{\overline{L}_0}\, \overline{r}_+^{\overline{J}_0^{3,+}}\, \overline{r}_-^{\overline{J}_0^{3,-}}
  }\Big{|}_\mathrm{BTZ}
  \nonumber   \\
   & \=
  \Tr_{_\mathrm{NS}}
  \pqty{
  q_{_\mathrm{AdS}}^{L_0}\, r_{+, _\mathrm{AdS}}^{J_0^{3,+}}\,  r_{-, _\mathrm{AdS}}^{J_0^{3,-}}\, \overline{q}_{_\mathrm{AdS}}^{\overline{L}_0}\,  \overline{r}_{+, _\mathrm{AdS}}^{\overline{J}_0^{3,+}}\, \overline{r}_{-, _\mathrm{AdS}}^{\overline{J}_0^{3,-}}}  \Bigg{|}_\mathrm{AdS}
  \label{eq:btzNS}       \\
  \mathscr{Z}^\mathrm{R}_{_\mathrm{CFT}}(\tau, \rho_{+}, \rho_-, \overline{\tau},\overline{\rho}_+ , \overline{\rho}_-) \Big{|}_\mathrm{BTZ}
   & \=
  \Tr_{_\mathrm{R}}\pqty{q^{L_0}\, r_+^{J_0^{3,+}}\,  r_-^{J_0^{3,}}\, \overline{q}^{\overline{L}_0}\, \overline{r}_+^{\overline{J}_0^{3,+}}\, \overline{r}_-^{\overline{J}_0^{3,-}}}\Bigg{|}_\mathrm{BTZ} \nonumber \\
   & \=
  \Tr_{_\mathrm{NS}}\pqty{(-1)^{F_s}\, q_{_\mathrm{AdS}}^{L_0}\, r_{+, _\mathrm{AdS}}^{J_0^{3,+}}\,  r_{-, _\mathrm{AdS}}^{J_0^{3,-}}\, \overline{q}_{_\mathrm{AdS}}^{\overline{L}_0}\,  \overline{r}_{+, _\mathrm{AdS}}^{\overline{J}_0^{3,+}}\, \overline{r}_{-, _\mathrm{AdS}}^{\overline{J}_0^{3,-}}}
  \Bigg{|}_\mathrm{AdS}\,.
  \label{eq:btzR}
  \end{align}
\end{subequations}
The parameters for the trace in the AdS geometry are given as
\begin{equation}\label{eq:taurhomapbtz}
q_{_\text{AdS}} \= e^{-\frac{2\pi i}{\tau }}  \,,
\qquad
(r_\pm)_{_\text{AdS}} \= e^{-2\pi i\,\frac{\rho_\pm}{\tau}}\,.
\end{equation}
We will analyze the two cases in turn. Note that in these traces we have dropped the zero-point energy, which we will include at the end (assuming that it is given by the semiclassical gravity answer). 

A note on our notation: in the rest of the section, we  use $\tau$,
$\rho_\pm$ to denote the modular parameter and chemical potentials in the BTZ geometry. In particular, when we attempt to understand the near-extremal limit we set
\begin{equation}\label{eq:TOmega}
T = \frac{1}{i\pi}\, \frac{1}{\overline{\tau} - \tau} \,, \qquad \Omega = \pi
(\tau + \overline{\tau}) \,,
\end{equation}
and take the limit
\begin{equation}\label{eq:nextlim}
\tau \to i\,\infty\,, \qquad \overline{\tau} \to - i\,0\,,\qquad
\text{holding}\;\ \abs{\tau} < 1 \,.
\end{equation}
The corresponding data for the \AdS{3} geometry will be indicated with an
explicit subscript as in~\eqref{eq:taurhomapbtz}. 

Below we find corrections from the one-loop determinant at temperatures of order \hbox{$T_\mathrm{gap} \sim \frac{1}{c}$}. It is useful to  define an energy scale $E_\mathrm{gap} = \frac{3}{c}$, inspired by the gap scale in the examples with the small 
$\mathcal{N}=4$ algebra~\cite{Heydeman:2020hhw}. 
To deal with factors in the intermediate calculations, it is convenient to define $T_\mathrm{gap} = \frac{4}{\pi^2}\,E_\mathrm{gap}$. 
It is important to emphasize that we take the low-temperature limit after sending $c\to \infty$. Therefore, our results are valid in the regime 
$T_\mathrm{gap}\, e^{-S_\mathrm{tree}} \ll T \ll T_\mathrm{gap}$, viz., we are insensitive to non-perturbative corrections scaling as~$e^{-c}$.

\subsection{The NS sector trace around BTZ}\label{sec:nsBTZ}

In this sector we are not preserving any supersymmetry, so there are no fermion zero modes. However, the presence of conserved currents implies that there should be bosonic zero modes. In fact, we expect from a semiclassical supergravity analysis a set of $10$ bosonic zero modes, $3$ from $\mathrm{SL}(2,\mathbb{R})$ isometry, $3$ from each of the $\mathrm{SU}(2)$ R-symmetries, and an addition one from the $\mathrm{U}(1)$ isometry.

The details can be worked out starting from our result for the thermal \AdS{3}
partition sum with antiperiodic fermion boundary conditions. We have
\begin{equation}
\begin{split}
\Tr_{_\mathrm{NS}}\pqty{q_{_\mathrm{AdS}}^{L_0}\, r_{+, _\mathrm{AdS}}^{J_0^{3,+}}\,
r_{-, _\mathrm{AdS}}^{J_0^{3,-}}\,
\overline{q}_{_\mathrm{AdS}}^{\overline{L}_0}\,  \overline{r}_{+,
  _\mathrm{AdS}}^{\overline{J}_0^{3,+}}\, \overline{r}_{-,
  _\mathrm{AdS}}^{\overline{J}_0^{3,-}}}  \Bigg{|}_\mathrm{AdS}
\= e^{S_\text{tree}} \times \abs{\schr{\text{NS}}^\gamma(q,r_+, r_-; 0)}^2 \times
\cdots \,.
\end{split}
\end{equation}
We have indicated here only the contribution from the spacetime supercurrents,
with the ellipsis indicating the contribution from states with non-vanishing
twist. This analysis is valid as long as $k>1$.

We would now like to understand the implications for the result in the
near-extremal limit of the BTZ geometry by taking the limit~\eqref{eq:nextlim}.
The main contributions come from the holomorphic part of the character, which
can be straightforwardly analyzed using the modular properties of the Jacobi
theta functions, and working out the limiting behavior of the sums in the
function $\mu(q,r_+,r_-\,;0)$. We find\footnote{ To arrive at the result quoted, we only performed a modular transformation of the character $\schr{\text{NS}}^\gamma(q,r_+, r_-; 0)$. The factors proportional to $T\,\rho_\pm^2$ originate from this manipulation. We should note these factors are absent when one considers the path integral computation of the trace~\cite{Kraus:2006nb}. This observation has been exploited in extracting the microcanonical density of states of supersymmetric black holes~\cite{Heydeman:2020hhw}. We, however, will leave the factor in place in our formulae for clarity, since our analysis of the spectral flow sum~\eqref{eq:sflowsum} does not completely justify it.~\label{fn:gravanomaly}  }
\begin{equation}
\mathscr{Z}^\mathrm{NS}_{_\mathrm{CFT}}(\tau,\rho_\pm, \overline{\tau},
\overline{\rho}_\pm) \bigg|_\mathrm{BTZ}
 \, \sim\,
e^{S_\mathrm{tree} + \beta+ \frac{T}{\gamma\, T_\mathrm{gap}}\, \left( 
    \rho_+^2 + \alpha\, \rho_-^2\right)}\, \pqty{\frac{T}{T_\mathrm{gap}}}^5\,  \mathfrak{F}_+(\rho_+)\,
\mathfrak{F}_-(\rho_-) \,,
\end{equation}
with
\begin{equation}
\mathfrak{F}_\pm(\rho)
 \;\equiv\;
\sum_{m\in \mathbb{Z}}\, \frac{\pqty{m + \frac{\rho}{2}}}{\sin(\pi\, \rho)}\;
e^{-4\,\frac{\mathsf{k}_\pm}{\mathsf{k}}\,\frac{T}{T_\mathrm{gap}}\,   \pqty{m + \frac{\rho}{2}}^2 }\,.    
\end{equation}
This is the desired answer in the grand canonical ensemble around a single BTZ
saddle with specified chemical potentials for the $\sut$ R-charges. Note that
a factor of $T^2$ comes from the modular transform of the theta functions,
while the sums defining the function $\mu_{_\mathrm{NS}}(q,r_+,r_-;0)$ give three additional 
powers of $T$ in the low-temperature expansion. There is also a non-vanishing zero-point energy contribution because the fermions are anti-periodic around the thermal circle. 

We can convert the result to fixed $\sut$ charge sector by Poisson
resumming the functions $\mathfrak{F}_\pm(\rho)$. The sum in the form given
can be discerned to be the derivative of $\vartheta_3(\rho,\tau)$ with respect to
the elliptic parameter $\rho$. The calculation is straightforward, and we simply
record the final result,
\begin{equation}
\mathfrak{F}_\pm(\rho) = 
\frac{\pi^\frac{3}{2}}{4}\, \pqty{\frac{\mathsf{k}}{\mathsf{k}_\pm} \,\frac{T_\mathrm{gap}}{T}}^{\frac{3}{2}}\,
\sum_{n=1}^\infty\, n\, \chi_\frac{n-1}{2}(r)\,
e^{- \frac{\pi^2}{4}\, \frac{\mathsf{k}}{\mathsf{k}_\pm}\, \frac{T_\mathrm{gap}}{T}\, n^2}\,.
\end{equation}
Note that we have lost a factor $T^{\frac{3}{2}}$ in the process. With this rewriting, we can project onto the fixed $\mathrm{SU}(2)$ charged
sector for each~$\sut$. One indeed expects no contribution from the $\mathrm{SU}(2)$ zero
modes once we fix the charges.

We did not turn on a chemical potential for the $\mathrm{U}(1)$ symmetry. Had we done so, a similar analysis would have indicated that the $\mathrm{U}(1)$ modes drop out since we would get a factor of $T^{-\frac{1}{2}}$, cf.~\cite{Mertens:2019tcm}. Therefore, we would have realized that in the ensemble with all the
charges fixed, the low-temperature behavior is given by $T^\frac{3}{2}$. All told, the partition function in the canonical ensemble with fixed charges fixed will behave as
\begin{equation}
\begin{split}
\widetilde{\mathscr{Z}}^{\mathrm{NS}}_{_\mathrm{CFT}} \pqty{T,\rho_+, \rho_-,
  Q_{\mathrm{U}(1)}=0 }\bigg|_\mathrm{BTZ}
 & \; \sim  \;
 \frac{\alpha^\frac{3}{2}}{(1+\alpha)^3}
\pqty{\frac{T}{T_\mathrm{gap}}}^\frac{3}{2}\,          
e^{S_\mathrm{tree} +\, \beta \,+ \frac{\rho_+^2 +\, \alpha\, \rho_-^2}{\gamma}\, \frac{T}{T_\mathrm{gap}} } \\ 
& \quad \times \sum_{m,n =\,1}^\infty\, m\,n\, 
\chi_{\frac{m-1}{2}}(r_+)\, \chi_{\frac{n-1}{2}}(r_-)\,
e^{-\beta E_\mathrm{gap} \frac{\alpha\,m^2 + n^2}{1+\alpha}}
\,.
\end{split}
\end{equation}
We wrote the answer in terms of  an energy gap scale, which we defined to be $E_\mathrm{gap} = \frac{3}{c}$. The leading low-temperature answer comes from states with non-zero charges     under the $\sut_\pm$ R-symmetries. The partition function therefore behaves as 
\begin{equation}\label{eq:NSfinal}
\begin{split}
\widetilde{\mathscr{Z}}^{\mathrm{NS}}_{_\mathrm{CFT}} \pqty{T,j_+, j_-,
  Q_{\mathrm{U}(1)}=0 }\bigg|_\mathrm{BTZ}
 & \sim 
\pqty{\frac{T}{T_\mathrm{gap}}}^\frac{3}{2}\,          
e^{S_\mathrm{tree} + S_1\, T -
\beta \,\frac{E_\mathrm{gap}}{1+\alpha}\,\pqty{ \alpha\, (2\,j_++1) ^2 +
(2\,j_-+1)^2 + E_1 } } \,.
\end{split}
\end{equation}
Here $S_1$ is the classical correction to the extremal black hole's thermodynamic entropy, and $E_1$ is the shift in the 
ground state energy (which includes the contribution from the zero-point energies). 
This is the result for fixed $\sut_\pm$ spins $j_\pm$, and we have indicated  the leading classical corrections to the entropy and the energy above extremality.
The scaling of the partition function as $T^\frac{3}{2}$ is as predicted from the behavior of the Virasoro vacuum character~\cite{Ghosh:2019rcj,Pal:2023cgk}. 
This temperature dependent pre-factor leads to a continuum density of states in the microcanonical ensemble. 

\subsection{The R sector trace around BTZ}\label{sec:rBTZ}

The trace in the Ramond sector around the BTZ solution is obtained from the thermal AdS answer with periodic fermion boundary conditions. 
Clearly, we will have a different zero mode structure owing to this change. 
We first analyze the behavior at a generic value of the $\sut_\pm$ chemical potentials. 
Subsequently, in~\cref{sec:irBTZ} we set them to appropriate values to insert a factor of  $(-1)^{F_s}$ into the R sector trace.

We therefore start with the thermal  \AdS{3} partition sum  with periodic fermion boundary conditions, which assuming a twist gap (for $k>1$) simplifies to the vacuum module contribution, viz., 
\begin{equation}
\begin{split}
\Tr_{_\mathrm{NS}}\pqty{(-1)^{F_s}\, q_{_\mathrm{AdS}}^{L_0}\, r_{+, _\mathrm{AdS}}^{J_0^{3,+}}\,
r_{-, _\mathrm{AdS}}^{J_0^{3,-}}\,
\overline{q}_{_\mathrm{AdS}}^{\overline{L}_0}\,  
\overline{r}_{+,_\mathrm{AdS}}^{\overline{J}_0^{3,+}}\, 
\overline{r}_{-,_\mathrm{AdS}}^{\overline{J}_0^{3,-}}}  
\Bigg{|}_\mathrm{AdS}
= e^{S_\text{tree}} \times \abs{\schr{\text{NS}}^\gamma(q,r_+, r_-; 1)}^2 \times
\cdots \,.
\end{split}
\end{equation}
Implementing the S-transform and taking the low-temperature limit, the right-movers give the exponential degeneracy through a non-vanishing tree-level contribution $S_\mathrm{tree}$. The left-movers get the dominant low-temperature contribution from the one-loop determinant factors and also from the spectral flow sum. Combining the pieces together, the  BTZ partition function may be expressed as 
\begin{equation}\label{eq:BTZRgen}
\begin{split}
\mathscr{Z}^\mathrm{R}_{_\mathrm{CFT}}(\tau,\rho_\pm, \overline{\tau},
\overline{\rho}_\pm) \bigg|_\mathrm{BTZ}
 & \;\sim\;
64\,\frac{T}{T_\mathrm{gap}}\, \exp\pqty{S_\mathrm{tree} + 2\,y\, 
  \pqty{ \,\rho_+^2 
+ \, \alpha\,\rho_-^2} }  
\\
&\qquad \times\; 
\frac{\cos^2\pqty{\frac{\pi\,(\rho_+ + \rho_-)}{2}}\,
\cos^2\pqty{\frac{\pi\,(\rho_+ - \rho_-)}{2}}}{\sin(\pi\,\rho_+)\, \sin(\pi\,\rho_-)} \, 
\mathfrak{B}_\alpha\pqty{y,\,\rho_+,\rho_-} \,.
\end{split}
\end{equation}
Here we have introduced the parameter
\begin{equation}\label{eq:ydef}
y 
\= \frac{1}{2}\, \frac{T}{T_\mathrm{gap}} \, \frac{\mathsf{k}_+}{\mathsf{k}} 
\= \frac{T}{2\,\gamma\, T_\mathrm{gap}}\,,
\end{equation}
which is the temperature in units of the gap temperature with some factors stripped off for simplicity.    
The function $\mathfrak{B}_\alpha$ is defined as a lattice sum 
\begin{equation}\label{eq:Bfndef}
\mathfrak{B}_\alpha(y,\rho_+,\rho_-)
\;\equiv\; 
\sum_{n_\pm \in \mathbb{Z}}\, 
\frac{\pqty{n_+ + \frac{1}{2}\,\rho_+}\,\pqty{n_- + \frac{1}{2}\,\rho_-}\ 
  e^{-8\,y\,
      \left(
\pqty{n_+ + \frac{1}{2}\,\rho_+}^2 + \alpha \,
\pqty{n_- + \frac{1}{2}\,\rho_-}^2 \right)}}{
  \bqty{1-4\,(n_+ + n_- + \frac{\rho_+ +\rho_-}{2} )^2 } \,
  \bqty{1-4\,(n_+ - n_- + \frac{\rho_+ - \rho_-}{2} )^2 }
}\,.
\end{equation}
This is the desired answer in the grand canonical ensemble around a single BTZ
saddle with specified chemical potentials for the $\sut$ R-charges.

To understand the low-temperature regime, one needs to estimate the sum. By 
redefining the summation variables $n_\pm$ and rotating the chemical potentials as   
\begin{equation}\label{eq:diagonalnm}
m_1 = n_+ + n_- \,, \qquad m_2 = n_+ - n_-\,, \qquad \rho_1 = \rho_+  + \rho_-\,,
\qquad
\rho_2 = \rho_+ - \rho_-\,,
\end{equation}
we first simplify $\mathfrak{B}_\alpha$  to  
\begin{equation}\label{eq:Balphadef}
\begin{split}
\mathfrak{B}_\alpha(y,\rho_1, \rho_2)
&\= 
\frac{1}{16}\, \sum_{m_1,m_2 \,\in \,\mathbb{Z}}\,  
     \bqty{ \frac{1}{1-4\,(m_1 + \frac{\rho_1}{2} )^2 } - 
      \frac{1}{1-4\,(m_2 + \frac{\rho_2}{2} )^2 }} \\
& \qquad \qquad \qquad  \times 
e^{- 2\,y
      \left(\pqty{m_1+m_2 + \frac{\rho_1+\rho_2}{2}}^2 +\, \alpha\, \pqty{m_1 - m_2 
      + \frac{\rho_1 - \rho_2}{2}}^2 \right)} \,.
\end{split}
\end{equation}

Let us start by considering generic values of $\rho_1$ and $\rho_2$. The sums over $m_1$ and $m_2$, which comprise a Gaussian lattice sum, can be completed in terms of a Riemann theta function. 
We then Poisson resum this lattice sum, which can be done straightforwardly as we explain in~\cref{sec:Rsums}.
The final result is the expression for $\mathfrak{B}_\alpha$ given in~\eqref{eq:Balphaval}. Using this, we end up with the  following expression for the R-sector partition function
\begin{equation}\label{eq:ZRfinal}
\begin{split}
\mathscr{Z}^\mathrm{R}_{_\mathrm{CFT}}(\tau,\rho_\pm, \overline{\tau},
\overline{\rho}_\pm) \bigg|_\mathrm{BTZ}  
& \,\sim\, 
\frac{4\,\sqrt{\alpha}}{1+\alpha}\,  
\pqty{\frac{T}{T_\mathrm{gap}}}^\frac{1}{2}\, \exp\pqty{S_\mathrm{tree} + 2\,y\, 
  \pqty{ \,\rho_+^2 
+ \, \alpha\,\rho_-^2 -\gamma} }  \\
&  \times  
\cos^2\pqty{\frac{\pi\,(\rho_+ + \rho_-)}{2}}\,
\cos^2\pqty{\frac{\pi\,(\rho_+ - \rho_-)}{2}}  \;  \int_0^{2\,\gamma\,y}\, du\, \frac{e^{u}}{\sqrt{u}} \, \mathfrak{f}_\alpha (u)\,.
\end{split}    
\end{equation}
The integrand $\mathfrak{f}_\alpha(u)$ has an expansion in terms of $\sut_\pm$ characters (as in the NS sector) and is given by 
\begin{equation}
\begin{split}
 \mathfrak{f}_\alpha (u)
 &\= 
\sum_{m_1,m_2=0}^\infty\, \pqty{1-\frac{\delta_{m_1,0}}{2}}\, \pqty{1-\frac{\delta_{m_2,0}}{2}}\, \mathrm{sign}(m_1-m_2) \, e^{- \frac{\pi^2\,m_2^2}{2\,y\,(1+\alpha)} } \\ 
& \qquad \qquad \times \bqty{
e^{-\frac{\pi^2\, (m_1 + \eta\,m_2)^2}{4\,u} } \, 
\chi_{\frac{m_1+m_2-1}{2}}(\rho_+)
\, \chi_{\frac{\abs{m_1-m_2}-1}{2}}(\rho_-) + (m_2 \leftrightarrow -m_2)}.
\end{split}
\end{equation}
In this expression, we introduced  
\begin{equation}\label{eq:etadef}
\eta \= 1-2\, \gamma  \=  \frac{1-\alpha}{1+\alpha}   
\end{equation}
for notational simplicity. 

The density of states in the Ramond sector can be read off now directly from~\eqref{eq:ZRfinal}. The answer is already expressed in terms of $\sut_\pm$ characters, which allows us to isolate the contribution from specific charge sectors. By taking a Laplace transform, 
we can calculate the microcanonical density of states. 
We now extract some general features without implementing the Laplace transform explicitly.

Firstly, the prefactor of $\sqrt{T}$ will disappear when we work with fixed $\mathrm{U}(1)$ charges. 
Secondly, the trigonometric factors involving $\rho_\pm$ encode the contribution of the fermionic zero modes in the background. 
As noted at the outset, this is a feature that distinguishes the NS sector trace from the R sector one. 
Finally, the integral of $\mathfrak{f}_\alpha$ contains non-trivial information. 
We can estimate this quantity -- the leading contribution comes from the terms with $m_1 = 1$ and $m_2 =0$ (or vice versa).  Thus, we arrive at our prediction for the leading contribution to the Ramond sector partition function at fixed $\mathrm{U}(1)$ charge, 
\begin{equation}
\begin{split}
\widetilde{\mathscr{Z}}^\mathrm{R}_{_\mathrm{CFT}}(T, \rho_+,\rho_-, Q_{\mathrm{U}(1)} = 0) \bigg|_\mathrm{BTZ} 
& \,\sim\, 
    \cos^2\pqty{\frac{\pi\,(\rho_+ + \rho_-)}{2}}\,
    \cos^2\pqty{\frac{\pi\,(\rho_+ - \rho_-)}{2}}\,
     \\
& \qquad \quad \times 
   \pqty{\frac{T}{T_\mathrm{gap}}}^\frac{3}{2}\,
   e^{S_\mathrm{tree} + \frac{T}{\gamma\, T_\mathrm{gap}} 
  \pqty{ \,\rho_+^2 + \, \alpha\,\rho_-^2 -\gamma} -  \beta\,E_\mathrm{gap}}\,.
\end{split}
\end{equation}
The power-law prefactor is the familiar $T^\frac{3}{2}$ factor that is expected from the near-horizon Schwarzian mode. The exponent captures the spectral gap, which we have conveniently parameterized in terms of the scale $E_\mathrm{gap}$. 
The main point to realize is that the partition function is exponentially suppressed at low temperatures, analogously to the NS sector trace~\eqref{eq:NSfinal}. Therefore, generically, the Ramond sector trace is exponentially suppressed in the grand canonical ensemble. A Laplace transform to the microcanonical ensemble will result in a continuum density of states, owing to the polynomial pre-factor $\sim T^\frac{3}{2}$. The technical details are analogous to the NS sector case, or to the general discussion of non-BPS near-extremal black holes~\cite{Iliesiu:2020qvm}.

\subsection{The index for R sector states}\label{sec:irBTZ}

In~\cref{sec:BPSN4} we saw that the~$\mathcal{N}=2$ index that accounts for the BPS states in the NS sector is given by setting~$\rho_{+,\text{AdS}}=\gamma \,\tau$, 
$\rho_{-,\text{AdS}}=-(1-\gamma)\,\tau$. As noted there, we could consider a similar trace weighted by fermion number in the R-sector. 
In the BTZ variables we are currently discussing, we obtain the special values $\rho_{+}=\gamma$, $\rho_{-}= -\, (1-\gamma)$. Equivalently, in terms of the rotated potentials introduced in~\cref{eq:diagonalnm} one needs~$\rho_1=2\, \gamma-1 = -\eta$ and $\rho_2=1$.  We will attack this computation by starting with the R-sector trace examined in~\cref{sec:rBTZ} and specialize to this choice value of the chemical potentials. From the string worldsheet perspective, this is the best we can do, since we have not quantized the string directly in the BTZ background.  

We will find it convenient to first consider setting $\rho_2=1$ and leaving $\rho_1$ arbitrary. The Hilbert space trace we consider is therefore (dropping the antiholomorphic arguments for notational simplicity)
\begin{equation}\label{eq:BTZRindexA}
\mathscr{Z}^\mathrm{R}_{_\mathrm{CFT}}(\tau,\rho_+=1 + \rho_-) \bigg|_\mathrm{BTZ}
 \=
\Tr_{_\mathrm{R}}\pqty{q^{L_0}\, e^{i\pi\, (J_0^{3,+} - J_0^{3,-})}\, e^{i\pi\, \rho_1\,(J_0^{3,+} + J_0^{3,-}) }}\Bigg{|}_\mathrm{BTZ}\,.
\end{equation}
At these special values, note that the partition function has a double zero from the trigonometric prefactor in~\eqref{eq:BTZRgen}, which as~$\rho_2 \to 1$ behaves as
\begin{equation}
 \frac{\cos^2\pqty{\frac{\pi}{2}\,\rho_1}\, \cos^2\pqty{\frac{\pi}{2}\,\rho_2}}
    {\sin\pqty{\frac{\pi}{2}\,(\rho_1+\rho_2)}\,
    \sin\pqty{\frac{\pi}{2}\,(\rho_1-\rho_2)}} 
   \,\to\,  - \frac{\pi^2}{4}\, (1-\rho_2)^2   + \order{(\rho_2 -1)^3}\,.
\end{equation}
Further, the sum~\eqref{eq:Balphadef} defining the function $\mathfrak{B}_\alpha$ has a simple pole at~$\rho_2=1$. This arises from 
the summand with $m_2 =0$ or $m_2 =-1$ and arbitrary $m_1$,
and cancels one of the zeros, as we discuss below.
The second zero is interpreted as originating from the fermionic states in the large $\mathcal{A}_\gamma$ algebra, as discussed in~\cref{sec:BPSN4}. 
They can be absorbed by the insertion of a fermion bilinear, as in~\cite{Gukov:2004fh}. Equivalently, this additional piece would not be present in the $\widetilde{\mathcal{A}}_\gamma$ character. 

Consider therefore computing the limit $\rho_2 \to 1$,  keeping $\rho_1$ arbitrary for the moment. For simplicity, we will examine the following function
\begin{equation}
 \mathfrak{C}_\alpha(y,\rho_1) \= \lim_{\rho_2 \to 1}\, 
  \cos(\frac{\pi}{2}\,\rho_2) \, \mathfrak{B}_\alpha(y,\rho_1,\rho_2) \,.
\end{equation}
The zero of~$\cos(\frac{\pi}{2}\,\rho_2)$ in the limit
kills all terms except for 
the terms with $m_2 =0$ and $m_2 =-1$ in the second term in the square brackets of~\eqref{eq:Balphadef}. 
Their coefficients are convergent sums in $m_1$.  
To wit,   
\begin{equation} \label{eq:defCalpha}
\begin{split}
\mathfrak{C}_\alpha(y,\rho_1)
&= 
-\frac{\pi}{64}\, e^{-\frac{2\,\alpha}{1+\alpha}\,y}\, 
\sum_{m\in \mathbb{Z}}\, \bqty{
e^{-2\, y\, (1+\alpha) \,
\pqty{m + \frac{1}{2}\, \rho_1 + \frac{1}{2}\,\eta}^2 } 
-e^{-2\, y\, (1+\alpha) \,
\pqty{m + \frac{1}{2}\, \rho_1 - \frac{1}{2}\,\eta}^2 } }\\
&=
-\frac{\pi^{\frac{3}{2}}\, e^{-2\,\gamma\,y}}{64\,\sqrt{2\,y\,(1+\alpha)}}\,
 \bqty{
\vartheta_3\pqty{e^{i\,\pi\, \pqty{\rho_1 +\eta}}, e^{-\frac{\pi^2}{(1+\alpha)\,y}}} - \vartheta_3\pqty{e^{i\,\pi\, \pqty{\rho_1 -\eta}}, e^{-\frac{\pi^2}{(1+\alpha)\,y}}}} \\ 
&= 
\frac{\pi^{\frac{3}{2}}\, e^{-2\,\gamma\,y}}{16\, \sqrt{2\,y\,(1+\alpha)}}\, 
\sum_{n=1}^\infty\, \sin(n \pi\, \rho_1)\, \sin\pqty{n\,\pi\,\eta}\; 
e^{-\frac{\pi^2}{2\,y\,(1+\alpha)}\, n^2} \,.
\end{split}
\end{equation}
One recognizes the sum in the first line as defining the theta function, and the second line is  obtained by a modular transformation (or equivalently a Poisson summation). 
We have expressed the result in terms of the parameter $\eta$ defined in~\eqref{eq:etadef}. 
 
Using the answer for the sum, we learn that the  partition function has the following low-temperature behavior  
\begin{equation}\label{eq:ZRindexA}
\begin{split}
\mathscr{Z}^\mathrm{R}_{_\mathrm{CFT}}(T,\rho_+=1 + \rho_-) \bigg|_\mathrm{BTZ} 
&\,\sim\, 
    2\,\pi^2\, \frac{\sqrt{\pi\,\alpha}}{1+\alpha}  \,(\rho_2-1)\, 
    \sqrt{\frac{T}{T_\mathrm{gap}}}\, \sin(\pi\rho_1)\,
    e^{S_\mathrm{tree} + S_1' \,\frac{T}{T_\mathrm{gap}}}  \,  \\ 
&\qquad   
    \times \sum_{n=1}^\infty\, \chi_{\frac{n-1}{2}}(\rho_1) \, \sin\pqty{n\pi\,\frac{1-\alpha}{1+\alpha}}\; 
e^{-\frac{\pi^2\,\alpha}{(1+\alpha)^2}\,\beta\, T_\mathrm{gap}\, n^2} \,.
\end{split}
\end{equation}
Here $S_1'$ is the coefficient of the term linear in $T$ in the exponent. It includes the chemical potential contribution from the first line of~\eqref{eq:BTZRgen} (cf.~\cref{fn:gravanomaly}), and the piece $e^{-2\,\gamma\,y}$ in $\mathfrak{C}(y,\rho_1)$. Explicitly, as a function of $\rho_1$ it is
\begin{equation}
S_1' =  -1 + \frac{(1+\alpha)^2}{4\,\alpha}\, \pqty{1 + \rho_1^2 + 2\,\rho_1\, \eta} \,.
\end{equation}
%

From~\eqref{eq:BTZRindexA} it is clear that $\rho_1$ is the chemical potential that couples to $\mathsf{J}_1^3 = \frac{1}{2} (J_0^{3,+} + J_0^{3,-})$. Therefore, we can interpret the above sum capturing the contribution of states in the representation with spin $\mathsf{j} = \frac{n-1}{2}$. The energy of these states can be read off from the exponent to be  
\begin{equation}\label{eq:Ejdef}
E_{_\mathrm{R}}(\mathsf{j}) 
\= 
\frac{16\,\alpha}{(1+\alpha)^2}\, \pqty{\mathsf{j} + \frac{1}{2} }^2\, E_\mathrm{gap}
\end{equation}
This is, in fact, the energy above the threshold value $c/24$ that we have seen R-sector states must have in~\eqref{eq:Runitc}.  Before proceeding further, let us fix the $\mathrm{U}(1)$ charge and soak up the fermion zero mode (i.e., remove the factor $(\rho_2-1 )\,\sqrt{T}$ from~\eqref{eq:ZRindexA}) and write the result as 
\begin{equation}\label{eq:ZRindexB}
\begin{split}
\widetilde{\mathscr{Z}}^\mathrm{R}_{_\mathrm{CFT}}(T,\rho_+=1 + \rho_-,
Q_{\mathrm{U}(1) =0}) \bigg|_\mathrm{BTZ} 
&\,\sim\, 
   2\,\pi^2\, \frac{\sqrt{\pi\,\alpha}}{1+\alpha}  \,
    \sin(\pi\rho_1)\,
    e^{S_\mathrm{tree} + S_1' \,\frac{T}{T_\mathrm{gap}}}  \,  \\ 
&  \qquad 
    \times \sum_{\mathsf{j}=0}^\infty\, \chi_{\mathsf{j}}(\rho_1) \, \sin\pqty{(2\,\mathsf{j} + 1)\,\pi\,\eta}\; 
e^{-\beta\, E_{_\mathrm{R}}(\mathsf{j})} \,.
\end{split}
\end{equation}
A notable feature here is the absence of a temperature dependent pre-factor. Therefore, one can immediately read-off the microcanonical density of states. 
One interesting feature is the sinusoidal modulation as a function of $\alpha$ 
in the number of states at fixed $\mathsf{j}$. 

For the $\mathcal{N}=2$ index computation described in~\cref{sec:BPSN4}, 
we want to further specify to $\rho_1  = 2\,\gamma-1 = \frac{\alpha-1}{1+\alpha}$, whence the linear correction in the exponent vanishes, $S_1' = 0$. 
Explicitly,
\begin{equation}\label{eq:ZRindexC}
\begin{split}
\widetilde{\mathscr{Z}}^\mathrm{R}_{_\mathrm{CFT}}(T,\rho_+=\gamma,
\rho_- = \gamma -1,
Q_{\mathrm{U}(1) =0}) \bigg|_\mathrm{BTZ} 
&\,\sim\, 
    \frac{2\,\pi^\frac{5}{2}\,\sqrt{\alpha}\, e^{S_\mathrm{tree}}}{(1+\alpha)} 
      \,  \sum_{\mathsf{j}=0}^\infty\,  \sin^2\pqty{(2\,\mathsf{j} + 1)\,\pi\,\eta}\; 
e^{-\beta\, E_{_\mathrm{R}}(\mathsf{j})} \,.
\end{split}
\end{equation}
This index has an explicit temperature dependence, which arises because the BPS states contributing to the index have an energy above the threshold given by~\eqref{eq:Runitc}. 
In particular, there are no states in the theory at threshold and hence there is no constant term in the index. 
While there is a large degeneracy from the right-movers captured by $S_\mathrm{tree}$, the left-moving BPS states tensored with these large number of states 
have non-vanishing energy and reside at the bottom of the continua discussed in~\cref{sec:rBTZ}. 
While the  index serves to isolate them from the continuum, one doesn't see a clean separation of the states that contribute to the macroscopic black hole entropy (see also comments in~\cref{sec:discuss}). 
Further, our analysis also predicts a vanishing result at  $\alpha =1 $. One can check this directly from the sum~\eqref{eq:Balphadef}.  
This point with $k_+ = k_-$ is special with an additional twisted algebra~\cite{Defever:1988um}. We have not investigated the implications of this observation in detail, which one ought to do in light of the special features of the $k=1$ theory. 

\section{Discussion}\label{sec:discuss}

In this paper we have analyzed strings on the $\asx{\stso}$ background, focusing on the 
computation of the one-loop worldsheet partition function. Our calculation reproduces the spectrum of supergravity states~\cite{Eberhardt:2017fsi}, including the non-chiral states, in the semiclassical limit. The single string free energy can be succinctly described in terms of characters of the superisometry algebra. The free multi-string Fock space sum then reproduces the superconformal characters of the asymptotic symmetry algebra. This calculation demonstrates that the spectrum of the worldsheet theory correctly captures the boundary supergraviton multiplet, which corresponds to chiral currents of the dual spacetime CFT. In addition, we also find current bilinears as in other \AdS{3} compactifications. Most of this analysis largely parallels the $\asx{T^4}$ compactification analyzed in~\cite{Ferko:2024uxi}. 

One advantage of the $\asx{\stso}$ background is that one can analyze the theory for the entire range  $\lads^2/\ell_s^2 = k \geq 1 $ directly in the RNS formalism. We examine the $k=1$ theory, which has been argued to correspond to the tensionless string. 
We find that the boundary gravitons merge into a continuum representation in this case, as was suggested earlier~\cite{Gaberdiel:2018rqv,Gaberdiel:2024dva}. 
In fact, the theory has no discrete states in the sector with zero units of $\slt$ spectral flow. 
However, contrary to earlier expectations, it does have discrete states at higher spectral flow levels. 
The states also do not appear to fit into higher spin representations. We argue, therefore, that the $k=1$ theory analyzed herein is distinct from the tensionless string. It is surprising (but not unprecedented, as noted in~\cref{sec:intro}) that there are different consistent worldsheet CFTs that could be characterized as describing string scale AdS spacetimes. 

We also used the worldsheet partition function to extract the low-temperature behavior of near-extremal BTZ black holes. We see the trace over the Neveu-Schwarz or Ramond Hilbert  spaces of the boundary CFT is exponentially damped at low temperatures in the fixed charge ensemble. One can also compute the trace over the Ramond states, weighted by the spacetime fermion number. By tuning the chemical potentials for the $\sut$ currents to an appropriate value, and removing additional fermionic zero modes,  we can compute the BPS index of the theory defined in~\cite{Gukov:2004fh}. This index gets contribution from the Ramond sector BPS states, and has an explicit dependence on temperature.

To appreciate this point better, consider the more familiar example of  $\asx{K3}$ compactification, which preserves the small $\mathcal{N}=4$ symmetry. 
BTZ black holes at threshold energy $\frac{c}{12}$ are BPS. 
For these states, the R-charge, which is characterized by a single $\sut$ spin quantum number $j$, lies in the window $-\frac{k}{2} \leq j \leq \frac{k}{2}$. One might have thought that the states with any allowed value  $j$ contributes to the BPS degeneracy. This is not true once one includes the quantum effects which are unsuppressed in the near-horizon region~\cite{Heydeman:2020hhw}. The BPS degeneracy is concentrated on states with $j=0$, while states with $j >0$ are gapped with a spin dependent energy gap, and lie at the bottom of a continuum distribution.

BTZ black holes in the $\asx{\stso}$ compactification should correspond to Ramond sector states of the large $\mathcal{N} = 4$ asymptotic symmetry algebra saturating the unitarity bound. 
However, there are no states at the threshold energy $\frac{c}{12}$ since  the unitarity bound forbids the presence of $\sut$ singlets.  
Instead, one has a spin-dependent gap in the energy. 
Consequently, when we compute the index, we isolate the contribution of the BPS states,  thereby obtaining a prediction for their spectral distribution. 
However, analogously to states with non-zero $\sut$ spin in the $\asx{K3}$ compactification these states are not isolated but lie at the bottom of the continuum density of states.  

It is interesting to compare the fate of the exponentially large number of states of supersymmetric black 
hole in this theory compared to extremal black holes in other theories. 
Recall that for non-supersymmetric extremal black holes, there is no mass gap and the $e^{S_\mathrm{tree}}$ states effectively merge into the continuum. 
For supersymmetric black holes in theories with smaller supersymmetry algebras (either in AdS$_3$ or in asymptotically flat space), 
there are~$e^{S_\mathrm{tree}}$ supersymmetric states  at threshold and there is a mass gap separating them from the non-supersymmetric continuum states. The index picks out precisely these states.
In the case of the AdS$_3$ theory with the large $\mathcal{N}=4$ algebra, there are no states at threshold and the~$e^{S_\mathrm{tree}}$ supersymmetric states are buried inside the 
non-supersymmetric continuum states, all starting after a mass gap from threshold. Nevertheless,
the index still picks out these supersymmetric states.

It would be interesting to better understand the $k=1$ theory obtained herein. The spectrum we have obtained appears naively not to show the symmetry enhancement due to the absence of higher spin currents. It would also be useful to understand whether one arrives at the same theory by working in the hybrid formalism for $k> 1$. Since the boundary gravitons merge with the continuum, one should analyze how this affects the construction of the asymptotic symmetry algebra from the worldsheet as in~\cite{Giveon:1998ns,deBoer:1998gyt,Kutasov:1999xu}. Furthermore, as noted above, the large $\mathcal{N} =4$ algebra has additional automorphisms for $k_+  = k_-$ or $\alpha =1$. While we have noted the vanishing of the index for this value of parameters, the physical implication of this statement for the $k=1$ theory remains to be better understood. 
For example, are there additional fermionic zero modes  which would explain the vanishing of the R-sector index?

\section*{Acknowledgements}
It is a pleasure to thank Lorenz Eberhardt, Matthias Gaberdiel, and Joaquin Turiaci for illuminating discussions and for sharing their results prior to publication. 
We would also like to thank Christian Ferko for collaboration on related topics.
We are grateful to Matthew Heydeman, Xiaoyi Shi, and Joaquin Turiaci for coordinating the submission of the results to the arXiv. 

We would like to thank KITP and the Aspen Center for Physics for hospitality during the course of this work. The former was supported by the grant NSF PHY-2309135 to the Kavli Institute for Theoretical Physics (KITP) and the latter by National Science Foundation grant PHY-2210452.
S.M.~acknowledges the support of the STFC grants ST/T000759/1,  ST/X000753/1. M.R.~is supported by U.S.~Department of Energy grant DE-SC0009999.
M.R.~would also like to acknowledge the hospitality of the workshop ``Holographic Duality and Models of Quantum Computation'' held at Tsinghua Southeast Asia Center in Bali, Indonesia (2024) during the course of this project.

\appendix

\section{The large \texorpdfstring{$\mathcal{N} =4$}{N=4} superconformal algebra}\label[appendix]{sec:N4algcomm}

We record here the large $\mathcal{N} =4$ superconformal algebra. It is generated by the Virasoro currents $L_m$, supercurrents $G^a_r$, two sets of $\sut$ current algebra generators $J^{\pm,i}_m$, a $\mathfrak{u}(1)$ current algebra generator $U_m$, and finally four fermionic generators $Q_r^a$.
The indices $\{m,n\}$ and $\{r,s\}$ are the usual grading indices of the
superconformal algebra. The former are integers, and the latter integers (in
Ramond) or half-integers (in Neveu-Schwarz).
Here $\{i,j,l\} \in \{1,2,3\}$ represent vector indices of the $\sut$ currents,
while $\{a,b\} \in \{0,1,2,3\}$ are bispinor type indices (inherited from $\mathrm{SO}(4)$). We will need a set of
$4\times 4$
matrices $A^{\pm i}_{ab}$, which are
\begin{equation}
A^{\pm i}_{ab} = \frac{1}{2} \pqty{\pm \delta_{ia}\, \delta_{b0} \mp
  \delta_{ib}\, \delta_{a0} + \epsilon_{iab} } \,.
\end{equation}

The commutation relations of the Virasoro generators is the usual one with
\begin{equation}
c = \frac{6\, \mathsf{k}_+\, \mathsf{k}_-}{\mathsf{k}_+ + \mathsf{k}_-}\,.
\end{equation}
The commutators of the rest of the generators with the $L_m$ can be specified by
giving the weights of the latter. The fermionic generators $Q^a_r$ have weight $\frac{1}{2}$ and the supercurrents $G^a_r$ weight $\frac{3}{2}$. The bosonic generators $J^{\pm,i}_m$ and $U_m$ have weight $1$. The rest of the commutators are given as
\begin{equation}
\begin{split}
\{G_r^a, G_s^b\}
 & =
\frac{c}{3}\, \delta^{ab}\, \pqty{r^2 - \frac{1}{4}} \delta_{r+s,0} + 2\, \delta^{ab}\, L_{r+s} \\
 & \qquad
+ 4\,i\, (r-s)\, \pqty{\gamma\, A_{ab}^{+i} \, J_{r+s}^{+,i} + (1-\gamma)\, A_{ab}^{-i}\, J_{r+s}^{-,i}} \\
\{Q^a_r, G^b_s\}
 & =
2\, A^{+i}_{ab}\, J^{+,i}_{r+s}
- 2\, A^{-i}_{ab}\, J^{-,i}_{r+s}
+ \delta^{ab}\,
U_{r +s}    \\
\{Q^a_r, Q^b_s\}
 & = \frac{\mathsf{k}_+ + \mathsf{k}_-}{2}\, \delta^{ab}\, \delta_{r+s,0} \\
[J^{\pm,i}_m,G_r^a]
 & =
i\, A_{ab}^{\pm i}\, G^b_{m+r} \mp \frac{2\,\mathsf{k}_\pm}{\mathsf{k}_+ + \mathsf{k}_-}\,
m\, A_{ab}^{\pm i}\, Q^b_{m+r} \\
[U_m,G^a_r]
 & = m\, Q^a_{m+r}  \\
[J^{\pm,i}_m,Q^a_r]
 & =
i\,A^{\pm i}_{ab}\, Q^b_{m+r} \\
[J_m^{\pm,i},J_n^{\pm,j}]
 & = i\, \epsilon^{ijl}\, J^{\pm,l}_{m+n} + \frac{\mathsf{k}_\pm}{2}\,m\, \delta^{ij}\,
\delta_{m+n,0} \\
[U_m,U_n]
 & =
\frac{\mathsf{k}_+ + \mathsf{k}_-}{2}\, m \, \delta_{m+n,0} \,.
\end{split}
\end{equation}
It is clear from these that the supercharges transform as $(\mathbf{2}, \mathbf{2}, \mathbf{1})$ under the global $\sut_+ \oplus \sut_- \oplus \mathfrak{u}(1)$ algebra.

The superisometry algebra $\Dta$ is generated by $\{L_0, L_{\pm 1}\}$,
$G^a_{\pm \frac{1}{2}}$ and $J_0^{\pm, i}$ in the NS sector. In this limit the
commutators of the $Q^a_r$ decouple, and $U_0$ is a central element.

As originally explained in~\cite{Goddard:1988wv} the algebra $\mathcal{A}_\gamma$ can be simplified by factoring out the
$\mathfrak{u}(1)$ current algebra generators $U_m$, and the four free fermions $Q^a_r$. One usually writes the resulting algebra as $\widetilde{A}_\gamma$, so that
\begin{equation}
\mathcal{A}_\gamma = \widetilde{\mathcal{A}}_\gamma \oplus \mathcal{A}_\mathrm{QU}
\end{equation}
The algebra $\widetilde{\mathcal{A}}_\gamma$ itself only has the Virasoro
generators, the superconformal generators, and the $\sut_\pm$ current algebra
generators.  It can be characterized by noting that
\begin{itemize}[wide,left=0pt]
  \item The levels of the $\sut_\pm$ current algebras are now reduced by one unit,
        viz.,
        \begin{equation}
        \widetilde{\mathsf{k}}_\pm = \mathsf{k}_\pm -1 \,.
        \end{equation}
  \item The central charge $\widetilde{c}$ is reduced by 3,
        \begin{equation}
        \widetilde{c} = c -3 = 3\, \frac{2\, \widetilde{\mathsf{k}}_+
          \, \widetilde{\mathsf{k}}_- + \widetilde{\mathsf{k}}_+ +
          \widetilde{\mathsf{k}}_-}{\widetilde{\mathsf{k}}_+
          + \widetilde{\mathsf{k}}_- +2}\,.
        \end{equation}
  \item The anti-commutators of the supercurrents is modified in two essential
        ways. The piece that involves the $\sut_\pm$ generators have their
        coefficients modified with
        \begin{equation}
        \gamma \to \frac{\mathsf{k}_- -1}{\mathsf{k}_+ + \mathsf{k}_-}\,,
        \qquad
        1-\gamma \to \frac{\mathsf{k}_+ -1}{\mathsf{k}_+ + \mathsf{k}_-}\,.
        \end{equation}
        Furthermore, instead of the central charge $\widetilde{c}$ we find
        \begin{equation}
        \widetilde{c}' = 6\,\frac{(\mathsf{k}_+ -1)\, (\mathsf{k}_--1)}{\mathsf{k}_+ +
          \mathsf{k}_-}\,,
        \end{equation}
        and there are non-linear terms involving products of the $\sut$ generators.
\end{itemize}
For reference, let us also note that the BPS bound for
$\widetilde{\mathcal{A}}_\gamma$ is similar to~\eqref{eq:N4uni} and
can be expressed as
\begin{equation}
h \;\geq\; \frac{1}{\widetilde{\mathsf{k}}_+ + \widetilde{\mathsf{k}}_- +2} \bqty{
  (\widetilde{\mathsf{k}}_+ + 1) \, \ell_- + (\widetilde{\mathsf{k}}_- + 1)\, \ell_+ + (\ell_+ - \ell_-)^2 }\,.
\end{equation}
Accounting for shift of the levels we see that this the same bound as in
$\mathcal{A}_\gamma$ for states carrying no $\mathfrak{u}(1)$ charge. The
non-linear term is more readily understood because of the current bilinears in
the anti-commutators of the supercurrents as indicated above.

\section{The partition function in the tensionless limit}\label[appendix]{sec:tensionless}

We briefly outline the result for the worldsheet partition function for $k_+ = k_- = 2$, and comment on extracting the spectrum. The single particle free energy computed from the worldsheet simplifies to~\eqref{eq:fPFintegralk1}. In obtaining this expression, the bosonic contributions from $\sut_{k_\pm -2}$ have been set to unity. 

Our goal is to explain  how to obtain the expression presented in~\eqref{eq:k1fpf}.  
We begin by noting that the fermionic oscillators have a straightforward  Fourier expansion in the worldsheet variable $\zs$ with non-negative coefficients. One way to see this is to use the  infinite 
product expansion and write 
\begin{equation}\label{eq:k1fermionosc}
\frac{\prod\limits_{\epsilon_\pm = \pm} \, \vartheta_1\left(\frac{\tau \,+\, \epsilon_+\,\rho_+ \,+\, \epsilon_-\, \rho_-}{2},\ts \right)}{\eta(\ts)^6}
\=
\zs^{\frac{1}{4}}\,q^{-1} \, 
\bqty{1-q + \schgS\left(\frac{1}{2},\frac{1}{2},\frac{1}{2}\right)}
\, \sum_{\hosc\,\geq\, 0}\, D_{\hosc}(q,r_+,r_-)\, \zs^{\hosc}\,.
\end{equation}
As written, $D_{\hosc}$ defined in~\eqref{eq:k1fermionosc} contains terms with negative powers of $q$; the coefficient of $\zs^\hosc$ is a finite Laurent polynomial in $q$. Focus on the contribution  from $D_{\hosc} \supset \zs^{\hosc}\, q^{\josc}$ with $\josc \in \mathbb{Z}$.\footnote{ The dependence on the chemical potentials $r_\pm$ can be conveniently assembled into $\sut$ characters, but will not be necessary for the arguments below.} We can easily bound $\hosc \in \mathbb{Z}_{\geq0}$ using the infinite sum representation of the theta-function,  
\begin{equation}
\vartheta_1\pqty{\frac{\tau}{2},\ts} 
\= 
\sum_{m \in \mathbb{Z} + \frac{1}{2}} \, (-1)^m\, \zs^{\frac{1}{2}\, m^2}\, q^{\frac{1}{2}\,m}\,.
\end{equation}
Each element of this sum has~$\Delta = 2 m^2$, and using the Cauchy-Schwarz inequality,  
we obtain 
\begin{equation}\label{eq:Doscbds}
   \hosc = \frac{1}{2}\,(\josc-1)^2 + x -\frac12 \,, \qquad x\geq 0 . 
\end{equation}
A similar statement holds for the right moving oscillators. 

The expansion of the $\slt$ oscillator contribution can be done using the following formula (due to Don Zagier, see Appendix A of~\cite{Ferko:2024uxi}, and see also~\cite{Ashok:2020dnc}):
\begin{equation}\label{eq:theta1exp}
\frac{\eta(\ts)^3}{\vartheta_1(\tau,\ts)}
= \sum_{n\in \mathbb{Z}} \, (-1)^{n} \,
\bqty{\sum_{r \geq n+w+1} (-1)^r\, \zs^{\frac{r\,(r+1)}{2}}}\, \zs^{-\frac{n\,(n+1)}{2}}\, q^{-n-\frac{1}{2}}\,.
\end{equation}
This is valid in the regime $ w \, < \, \frac{\Im(\tau)}{\Im(\ts)} \,<\, w+1$, for $w \in \mathbb{Z}$, and is the origin of the spectral flow.

With these expansions, we can rewrite using~\eqref{eq:I0def} the integrand of~\eqref{eq:fPFintegralk1} as 
\begin{equation}
\mathcal{I}_{_\mathrm{PF}}
\=
\abs{1-q}^2\, I_0  \;
  \abs{\widetilde{\sum} \; (-1)^{n+r}\,
\; D_{\hosc}\;  \zs^{\frac{1}{4} + \frac{r\,(r+1)}{2} -\frac{n\,(n+1)}{2} + \hosc + p_L^2} \; q^{-n-\frac{3}{2} }}^2\,,
\end{equation}
where $\widetilde{\sum}$ is a shorthand for the sums over  $n \in \mathbb{Z}$, $r\, \geq \, n + w + 1$, $\hosc \, \geq 0$, and $p_L \, \geq \,0$, respectively. We can now do the integrals over $\ts_1$ and $\ts_2$ to obtain
\begin{equation}\label{eq:fPFaB}
\begin{split}
f_{_\text{PF}}
 & \=
 \abs{1-q}^2\, I_0 \; \sum_{w=0}^\infty\, 
 \widehat{\sum} \;  \, q^{-n-\frac{3}{2}}\, \qb^{-\overline{n}-\frac{3}{2}}\, 
 D_{\hosc}\, \overline{D}_{\hosc}\, \mathfrak{J}_{k=1} \,,\\
\mathfrak{J}_{k=1}
 & \, \equiv \,
\int_{-\infty}^{\infty}\, \frac{\zeta\,d\zeta}{i\pi}\,
\frac{e^{2i\,\beta\,\zeta}}{\zeta^2 + \Delta_w^2} \bqty{
  e^{-\frac{2\beta}{w+1}\,(\zeta^2 + \Delta_w^2)} - e^{-\frac{2\beta}{w}\,(\zeta^2 + \Delta_w^2) }} \,, 
\end{split}
\end{equation}
with
\begin{equation}\label{eq:DeltawB}
\Delta_w^2 = \frac{1}{4} + 
\frac{r\,(r+1)}{2} - \frac{n\,(n+1)}{2} + \hosc + p_L^2 \,.
\end{equation}
To write this compactly we introduced the shorthand notation  
\begin{equation}
\widehat{\sum} =  \sum_{n,\, \overline{n}\,\in \,\mathbb{Z}} \, 
\sum_{r \, \geq\,  n+w+1} \, \sum_{\overline{r} \, \geq \, 
\overline{n} + w +1} \, 
(-1)^{n+r+\overline{n} + \overline{r}}\, \sum_{\hosc,\,\hoscb}\, \sum_{p_L,p_R}\; \delta_\text{LM} \,,
\end{equation}
to indicate the sum over various variables. 
The level matching constraint from the $\ts_1$ integral, which  is explicitly,  
\begin{equation}\label{eq:k1lmatch}
\delta_\text{LM} : \quad
\frac{r\,(r+1)}{2} -\frac{n\,(n+1)}{2} + \hosc + p_L^2 =
\frac{\overline{r}\,(\overline{r}+1)}{2} -\frac{\overline{n}\,(\overline{n}+1)}{2} + \hoscb + p_R^2\,,
\end{equation}
may be used to remove one of the summations. 

We now want to argue that the chiral states are all contained in the prefactor $I_0$. In particular, the integral in~\eqref{eq:fPFaB} does 
not lead to new chiral states either in the discrete or in the continuum spectrum.  The essential point is that the bound~\eqref{eq:Deltarange} given in~\eqref{eq:fdSS} 
which for $k=1$ is  
\begin{equation}\label{eq:k1deltarange}
    \frac{w}{2} < \Delta_w < \frac{w+1}{2} \,.
\end{equation}
This constrains the allowed values of spacetime CFT conformal dimensions. In addition, it also  bounds the twist of the spacetime CFT both in the discrete and continuous parts of the spectrum.  

The discrete part of the spectrum arises from the residues of the poles at $\zeta = i\, \Delta_w$ in~\eqref{eq:fPFaB}. This part of the spectrum comprises
\begin{equation}\label{eq:fdk1}
    f_{_\mathrm{PF,disc}} \;\supset\; 
    \abs{1-q}^2 \, I_0\, D_{\hosc}(q)\, 
     \overline{D}_{\hosc}(\qb)\,
     q^{-n-\frac{3}{2}}\, \qb^{-\overline{n} - \frac{3}{2}} \, 
     (q\, \qb)^{\Delta_w}\,.
\end{equation}
On the other hand, the continuum states are obtained after combining terms from two different spectral flow sectors. We take second term from~\eqref{eq:fPFaB} at some $w \geq 1$ and the first term 
from the analogous integrand with one less unit of spectral flow ($w\to w-1$). For both terms the contour of integration is shifted by redefining 
$\zeta \to s + i\,\frac{w}{2} $. This results in the  contribution 
\begin{equation}\label{eq:fpcontk1B}
\begin{split}
f_{_\mathrm{PF,cont}}
& 
\,\supset \, \abs{1-q}^2\, I_0\,	\sum_{w=1}^\infty\; \widehat{\sum}
D_{\hosc}(q)\, 
\overline{D}_{\hosc}(\qb)\,
q^{-n-\frac{3}{2}}\, \qb^{-\overline{n} - \frac{3}{2}} 
\\ 
&\quad \times 
\int\limits_{-\infty}^\infty\, \frac{ds}{i\pi} \; 
\left(s+\frac{i\,w}{2}\right) \,(q\,\qb)^{\frac{w}{4}+ \frac{s^2}{w}}
\left[ 
    \frac{(q\,\qb)^{\frac{1}{w}\,\Delta_{w-1}^2}  }{\left(s+\frac{i\,w}{2}\right)^2 + \Delta^2_{w-1} } - 
    \frac{(q\,\qb)^{\frac{1}{w}\,\Delta_{w}^2}}{ \left(s+\frac{i\,w}{2}\right)^2 +\Delta_w^2 }
\right] \,.
\end{split}
\end{equation}

With these definitions, the contribution to the spectrum can be summarized as follows: 
\begin{itemize}[wide, left=0pt]
\item In the discrete part of the spectrum, one has states with 
\begin{equation}\label{eq:k1hsdisc}
\begin{split}
\hs   
& \= 
-n +  \Delta_w  - \frac{3}{2} +\josc + E_0\,,
\\
\hsb 
& \= 
-\overline{n} +  \Delta_w  
- \frac{3}{2} +\joscb + \overline{E}_0\,,    
\end{split}
\end{equation}
Here $E_0$ arises from the factor $1-q +\schgS\left(\frac{1}{2},\frac{1}{2},\frac{1}{2}\right)$ in~\eqref{eq:k1fermionosc} and is bounded $0\leq E_0 \leq 2$ upon using~\eqref{eq:dtalowspin} .  

\item Similarly, in the continuum part of the spectrum, one has states, for~$w \ge 1$, with
\begin{equation}\label{eq:k1hscont}
\begin{split}
\hs  
& \= 
- n + \frac{w}{4} + \frac{1}{w} (s^2 + \Delta_{w-1}^2 ) - \frac{3}{2} +  \josc + E_0\,,
\\
\hsb  
& \= 
-\overline{n} +  \frac{w}{4} + \frac{1}{w} (s^2 + \Delta_{w-1}^2) 
- \frac{3}{2} + \joscb + \overline{E}_0\,.
\end{split}
\end{equation}
The expression for the spacetime energy at the bottom of the continuum in the $w^{\mathrm{th}}$ spectral flow sector agrees with that of a discrete state at the top edge of the unitarity bound in the $(w-1)^{\mathrm{st}}$ spectral flow sector, as noted in~\cite{Maldacena:2000hw}.  
\end{itemize}

We shall first demonstrate that $\hs \geq 0$. To do so, set  $r = 1+ w + n + v$ with $v \geq 0$
and solve for $n$ in terms of $\Delta_w$ using~\eqref{eq:DeltawB}. 
Plugging this  back into~\eqref{eq:k1hsdisc} and using the relation~\eqref{eq:Doscbds}, one can simplify the resulting expression to
\begin{equation}
\begin{split}
\hs 
& \= 
\frac{1}{v+w+1}\pqty{\frac{1}{2}\,\pqty{\josc+w+v}^2 + x + p_L^2 - \frac{1}{4}  } +  \pqty{1 - \frac{\Delta_w}{v+w+1}}\, \Delta_w + E_0 \\
& \; \geq \; 
-\frac{1}{4(v+w+1)}
 +  \pqty{1 - \frac{\Delta_w}{v+w+1}}\, \Delta_w  \\
& \; > \; 
\frac{w}{2} - \frac{1+w^2}{4\,(1+v+w)} \,.
\end{split}
\end{equation}
To obtain the bound, we used non-negativity of $x$, $p_L$, $E_0$, and the perfect square which immediately gives the second line. This quantitiy is a quadratic in $\Delta_w$ that is monotone increasing in the range~\eqref{eq:Deltarange}. The lower bound is strict and is obtained from the limit $\Delta_w\to\frac{w}{2}$ from above. 
For $w \geq 1$ this argument immediately shows that $\hs >0$ establishing  unitarity. Clearly, a similar argument applies for $\hsb$. Since $\hs$ is strictly bounded above $\frac{1}{4}$ for $w \geq 1$,  it also makes it clear that 
there is no possibility of obtaining chiral states from this part of the spectrum. 

To wrap up the discussion of discrete states, let us look at states with $w=0$. For simplicity start with $\hosc  = \josc =0$. Furthermore, assume that the $\mathbf{S}^1$ is a large radius circle, so that momentum modes on it are nearly continuous. Then, since $\Delta_0^2 
= n\, (v+1) + \frac{(v+1)(v+2)}{2} + \frac{1}{4} + p_L^2$, we can have solutions for $n \leq -2$ provided we can fine tune $p_L$ such that~\eqref{eq:Deltarange}  is upheld. Such discrete states exist in the spectrum, but at generic point on the $\mathbf{S}^1$ moduli space they do not lead to states with $\hs =0$. Turning on $\hosc$ and $\josc$ can be checked not to help in this regard (for one $\hosc$ is a non-negative integer) and so one has to ensure that there is an admissible value of $n$. 

We can do a similar calculation
for the continuous spectrum for~$w \geq 1$ 
by solving for~$n$ using \eqref{eq:DeltawB} with $w \to w-1$. 
This gives, at~$s=0$, for~$w \geq 1$, 
\begin{equation}  \label{eq:conthbnd}
\begin{split}
\hs 
& \= 
-n + \frac{w}{4}  + \frac{\Delta_{w-1}^2}{w}  - \frac{3}{2} +\josc + E_0 \\
& \= \frac{w}{4} + 
\frac{1}{v+w}\pqty{\frac{1}{2}\,\pqty{\josc+w+v-1}^2 + x + p_L^2 - \frac{1}{4}  } +  
 \frac{v \, \Delta_{w-1}^2}{w(v+w)} + E_0 \\
 & \; \geq \; \frac{w}{4} -
\frac{1}{4(v+w)} +  
 \frac{v \, (w-1)^2}{4w(v+w)} \; \geq \; 0  \,.
\end{split}
\end{equation}
This establishes that the states in the continuous part of the spectrum with~$w \ge 1$ also satisfy spacetime unitarity. 

Now, can finally ask when we have chiral states in the continuum spectrum, i.e., when is the inequality in~\eqref{eq:conthbnd} saturated. 
For this we need to explore allowed values of parameters more carefully. 
Let us write the spacetime dimensions in the continuous part of the spectrum by substituting in~\eqref{eq:DeltawB} into~\eqref{eq:k1hscont} to obtain,
\begin{equation}
\begin{split}
\hs 
& \= 
-n + \frac{w}{4}  + \frac{\Delta_{w-1}^2}{w}  - \frac{3}{2} +\josc + E_0\,,\\
& \= 
\frac{\hosc + p_L^2}{w} + (\josc -1) + v + \frac{1}{4\,w} \pqty{1+ 3\,w^2 + 2\,v\,(1+2\,n + v)} + E_0\,.
\end{split}
\end{equation}
Realize that we have contributions from $\hosc = \josc  = p_L =0$. 
For $w=1$ and $v=0$ we pick up only the zero mode contribution $E_0$. When combined with the right movers (level matching is trivial in this case), we end up with states which correspond to chiral currents. However, their spins are clearly bounded from above by $2$.

In the main text we quoted the analog of~\eqref{eq:fPFaB} as our final result in~\eqref{eq:k1fpf}. In presenting it, we have redefined the summation variables 
\begin{equation}\label{eq:ZmapMO}
\frac{r\,(r+1)}{2} -\frac{n\,(n+1)}{2} + \hosc \to N - \frac{w\,(w+1)}{2} \,,
\end{equation}
and absorbed the factors of $\abs{1-q}^2$ into defining the coefficient $\widetilde{\mathsf{Q}}_{N,\overline{N}}^w$. Apart from these cosmetic changes for notational consistency, the rest is as stated above. In particular, the claim that we do not have additional chiral states can be argued directly as done at the end of~\cref{sec:stringk1}. In any event, the above discussion should make it transparent that there is no possibility of generating chiral states from the continuum. 

\section{Estimating the lattice sums for Ramond sector trace}\label[appendix]{sec:Rsums}

In this appendix, we explain the estimation of the function $\mathfrak{B}_\alpha(y,\rho_+,\rho_-)$, defined in~\eqref{eq:Bfndef} which is relevant for the computation of the Ramond sector trace. 

We want to argue that this function is exponentially suppressed for generic
values of $\rho_1$ and $\rho_2$. To get a quick sense of this, it is useful to
examine the special case $\alpha =1$, where there is an enhanced symmetry, and hence some simplifications. Setting $\alpha = 1$, the cross-term mixing
$m_1$ and $m_2$ in the exponent cancels, and we get the following simple result: 
\begin{equation}
\mathfrak{B}_1(y,\rho_1, \rho_2)
\= \mathfrak{b}(y,\rho_2 )\, \widehat{\mathfrak{b}}(y,\rho_1)  - \mathfrak{b}(y,\rho_1)\, \widehat{\mathfrak{b}}(y,\rho_2) \,,
\end{equation}
with the functions $\mathfrak{b}$ and $\widehat{\mathfrak{b}}$ are defined as
\begin{equation}
\begin{split}
\mathfrak{b}(y,\rho) 
& \,\equiv\,
\sum_{m\in \mathbb{Z}}\, e^{-4\,y\,\pqty{m+ \frac{\rho}{2}}^2} \,, \\
\widehat{\mathfrak{b}}(y,\rho) 
&\,\equiv\, 
\sum_{m\in \mathbb{Z}}\, \frac{e^{-4\,y\,\pqty{m+ \frac{\rho}{2}}^2}}{1-4\, \pqty{m + \frac{\rho}{2}}^2}\,,
\end{split}
\end{equation}
respectively. Either by direct Poisson resummation, or by noting that the sum defines the theta function $\vartheta_3$, we can obtain the low-temperature behavior of $\mathfrak{b}$ to be 
\begin{equation}
\mathfrak{b}(y,\rho) \= 
\sqrt{\frac{\pi}{4\,y}}\, \sum_{n\in \mathbb{Z}}\, e^{i \pi\,n\,\rho} \,.
\end{equation}
On the other hand, $\widehat{\mathfrak{b}}(y,\rho)$ can itself expressed in terms of $\mathfrak{b}$ since 
\begin{equation}
e^{-y}\, \dv{y}(e^y\, \widehat{\mathfrak{b}}(y,\rho)) = \mathfrak{b}(y,\rho)\,.
\end{equation}
This can be integrated up using the boundary condition $\widehat{\mathfrak{b}}(0,\rho) =0$ (which can be verified independently). Using these results, one can immediately conclude that 
\begin{equation}
  \mathfrak{B}_1(y,\rho_+, \rho_-) 
  \sim \frac{\pi}{8}\, \sin(\pi \rho_+)\,
  \sin(\pi\,\rho_-)\,  e^{-\frac{\pi^2}{4\,y}} + \cdots\,.
\end{equation}
Note that we have written the final answer in terms of the original chemical
potentials, making clear that the sums organize into $\sut_\pm$ characters. The
terms we dropped are exponentially suppressed at small $y$, relative to the
leading order one that 
we have retained. 

More generally, one can estimate the sums by exploiting the integral identity 
\begin{equation}\label{eq:1xexp}
\frac{1}{x} = \int_C\, dt\, e^{-x\,t}  \,.
\end{equation}
The contour is along a half-line where the integral converges and depends on $\mathrm{sign}(x)$.  For concreteness, let $\rho_1, \rho_2 \in (1,2]$. Then factors in the first line of~\eqref{eq:Balphadef} are non-positive, the contour is therefore along the negative half-line.  Implementing this, we obtain 
\begin{equation}\label{eq:BFdef}
\mathfrak{B}_\alpha(y,\rho_1, \rho_2)
\= \mathfrak{F}_\alpha(y,\rho_1, \rho_2) - \mathfrak{F}_\alpha(y,\rho_2,\rho_1) \,,
\end{equation}
where 
\begin{equation}
\mathfrak{F}_\alpha(y,\rho_1, \rho_2)
\=
e^{-2\,y\,\pqty{\alpha\, \rho_-^2 + \rho_+^2}}\,
\int_C\,dt\,  e^{-t \,(1-\rho_1^2)}\,\sum_{m_1,m_2\in \mathbb{Z}}
\, e^{ -\vb{m}^T\cdot
  \vb{A}\cdot \vb{m} + 2\pi \,i\, \vb{z}^T\cdot \vb{m} }\,.
\end{equation}
We have written the result in a form that makes manifest that the integrand is 
a lattice sum defines a Riemann theta function. The two-dimensional vectors $\mathbf{m}$ 
and $\mathbf{z}$ and matrix $\mathbf{A}$ 
\begin{equation}\label{eq:GHterm1}
\begin{split}
\vb{m} 
&= 
\mqty(m_1 \\ m_2) \,, \qquad 
\vb{z} 
= 
\frac{i\,(1+\alpha)}{\pi}\, y\,
\mqty(\pqty{1-\frac{2\,t}{y\,(1+\alpha)}}\,\rho_1 +
\eta\, \rho_2 \\
\rho_2  + \eta\, \rho_1)\,, \\ 
\vb{A} 
&= 
  2\,y\,(1 + \alpha)\, \mqty(1 -\frac{2\,t}{y\,(1+\alpha)} & \eta\\
\eta & 1 )\,.
\end{split}
\end{equation}
 It is useful to realize from this rewriting is that $\mathfrak{B}(0,\rho_1, \rho_2)$ vanishes (which may also be checked independently). The parameter $\eta$ is defined in~\eqref{eq:etadef}.

We will not directly use the properties of this lattice theta function.  Instead, we will exploit the fact that the integral form allows us to implement Poisson resummation in a straightforward manner. Fourier transforming the integrand, we find 
\begin{equation}
\mathfrak{F}_\alpha(y,\rho_1,\rho_2)
\= 
\frac{e^{-2\,\gamma\,y}}{64\,\sqrt{2}\,\sqrt{y\,(1+\alpha)}}\,
\int_{C_u} \, du\, \frac{e^u}{\sqrt{u}}\, \widetilde{\mathfrak{f}}_\alpha \,, \qquad 
u \=   2\,\gamma\, y - t \,.
\end{equation}
The integrand is the double sum 
\begin{equation}
\widetilde{\mathfrak{f}}_\alpha 
\= 
\sum_{m_1,m_2 \in \mathbb{Z}}\,
e^{-\frac{\pi^2}{4\,u}\, (m_1 + \eta\,m_2)^2 - 
\frac{\pi^2}{2\,y\,(1+\alpha)}\,m_2^2 }
\bqty{e^{i\pi (m_1\,\rho_1 + m_2\, \rho_2)} -
e^{i\pi (m_2\,\rho_1 + m_1\, \rho_2)} } \,.
\end{equation}
From this expression it is clear that one can integrate over $u$ and extract the low-temperature behavior. In fact, one can do better, since the sum itself can be reorganized into $
\sut_\pm$ characters by suitably rewriting the sum. Combining terms across diagonal quadrants in the $\mathbb{Z}^2$ lattice, we end up with 
\begin{equation}
\begin{split}
\widetilde{\mathfrak{f}}_\alpha 
& \=
-4\, \sum_{m_1=0}^\infty\, \sum_{m_2=0}^\infty\, \pqty{1- \frac{\delta_{m_1,0}}{2}}\, \pqty{1- \frac{\delta_{m_2,0}}{2}} \, e^{- \frac{\pi^2\,m_2^2}{2\,y\,(1+\alpha)} }\; \widetilde{F}_\alpha(m_1,m_2) \,,\\
\widetilde{F}_\alpha(m_1,m_2) 
&\= 
 \Bigg[
e^{-\frac{\pi^2\,(m_1 + \eta\,m_2)^2}{4\,u}} \, \sin( \pi(m_1+m_2) \rho_+)\, \sin(\pi(m_1-m_2)\,\rho_-) \\ 
&\qquad  \qquad 
+ e^{-\frac{\pi^2\, (m_1 - \eta\,m_2)^2}{4\,u}} \, \sin( \pi(m_1-m_2) \rho_+)\, \sin(\pi(m_1+m_2)\,\rho_-) \Bigg] \,.
\end{split}
\end{equation}
Putting all the pieces together, we arrive at our final result 
\begin{equation}\label{eq:Balphaval}
\begin{split}
\mathfrak{B}_\alpha 
& \= 
-\frac{1}{16\,\sqrt{2}}\, \frac{e^{-2\,\gamma\,y}}{\sqrt{y\,(1+\alpha)}}
\sin(\pi\,\rho_+) \, \sin(\pi\, \rho_-) \, 
\, \int_0^{2\,\gamma\,y}\, du\, \frac{e^{u}}{\sqrt{u}} \, \mathfrak{f}_\alpha\\
\mathfrak{f}_\alpha 
&\= 
\sum_{m_1=0}^\infty\, \sum_{m_2=0}^\infty\, \pqty{1- \frac{\delta_{m_1,0}}{2}}\, \pqty{1- \frac{\delta_{m_2,0}}{2}}\; 
e^{- \frac{\pi^2\,m_2^2}{2\,y\,(1+\alpha)} } \,  F_\alpha(m_1,m_2) \\
F_\alpha(m_1,m_2)
&\= 
\mathrm{sign}(m_1-m_2)\,\Bigg[
e^{-\frac{\pi^2\, (m_1 + \eta\,m_2)^2}{4\,u} } \, 
\chi_{\frac{m_1+m_2-1}{2}}(\rho_+)
\, \chi_{\frac{\abs{m_1-m_2}-1}{2}}(\rho_-) \\ 
&\qquad\qquad  \qquad \qquad \quad 
+ e^{-\frac{\pi^2\, (m_1 - \eta\,m_2)^2}{4\,u} } \, 
\chi_{\frac{\abs{m_1-m_2}-1}{2}}(\rho_+)
\, \chi_{\frac{m_1+m_2-1}{2}}(\rho_-)\Bigg] .
\end{split}    
\end{equation}
This is the result used to arrive  at~\eqref{eq:ZRfinal}. In writing the expression above, we have rescaled the sum $\widetilde{\mathfrak{f}}_\alpha $ and introduced  $\mathfrak{f}_\alpha$, which allows for a clear interpretation of the spectrum.


\providecommand{\href}[2]{#2}\begingroup\raggedright\endgroup

\end{document}